\documentclass[fleqn,usenatbib]{mnras}
\usepackage{ae,aecompl}
\usepackage{graphicx}	% Including figure files
\usepackage{amsmath}	% Advanced maths commands
\usepackage{amssymb}	% Extra maths symbols
\usepackage{amsbsy}
\usepackage{float}
\usepackage{color}
\usepackage{txfonts}
\usepackage{times}

\usepackage{morefloats}
\usepackage{multirow}

\usepackage{mathrsfs,bm}
\newcommand{\bcdot}{\ensuremath{%
  \mathchoice%
   {\mskip\thinmuskip\lower0.2ex\hbox{\scalebox{1.5}{$\cdot$}}\mskip\thinmuskip}}%
   {\mskip\thinmuskip\lower0.2ex\hbox{\scalebox{1.5}{$\cdot$}}\mskip\thinmuskip}%        
   {\lower0.3ex\hbox{\scalebox{1.2}{$\cdot$}}}%  
   {\lower0.3ex\hbox{\scalebox{1.2}{$\cdot$}}}%
}
\newcommand{\Msunh}{\>h^{-1}\rm M_\odot}
\newcommand{\Msun}{\>{\rm M_\odot}}

\newcommand{\Mpch}{\>h^{-1}{\rm {Mpc}}}
\newcommand{\Mpc}{{\rm \ {Mpc}}}
\newcommand{\kpch}{\>h^{-1}{\rm {kpc}}}
\newcommand{\kpc}{{\rm \ {kpc}}}

\newcommand{\Mag}{ \ {\rm mag} \ {\rm arcsec^{-2}}}
\def \MF{{\rm M}_{\rm 500}}

\title[MOCK OBSERVATIONS OF GALAXIES IN SIMULATIONS]{The Importance of Mock Observations in Validating Galaxy Properties for Cosmological Simulations}
\author[Tang et al.]{Lin Tang$^{1,2}$\thanks{tanglin23@mail.sysu.edu.cn},  Weipeng Lin$^{1, 3}$\thanks{linweip5@mail.sysu.edu.cn}, Yang Wang$^{1, 3}$, N. R. Napolitano$^{1, 3}$
  \vspace*{0.2cm}  \\
  $^1$School of Physics and Astronomy, Sun Yat-sen University, DaXue Road 2, 519082, Zhuhai, China
   \vspace*{0.2cm}  \\
   $^2$Physics and Space Science College, China West Normal University, 1 ShiDa Road, 637002, Nanchong, P.R.China
   \vspace*{0.2cm}  \\
   $^3$CSST Science Center for the Guangdong-Hongkong-Macau Greater Bay Area, DaXue Road 2, 519082, Zhuhai, China
  }
\date{}
\begin{document}

\label{firstpage}
\pagerange{\pageref{firstpage}--\pageref{lastpage}}

\maketitle
%-------------------------------------------------------%

\begin{abstract}
The galaxy luminosity function and galaxy stellar mass function are fundamental statistics {in the testing of} galaxy formation models. 
Theoretical predictions based on cosmological simulations can deviate from observations, especially at {the} bright and faint ends.
In this case, the mismatch may come from missing physics, oversimplified or inaccurate model recipes, or inappropriate methods of extracting basic astrophysical quantities from simulations. 
The latter is a crucial aspect to consider to avoid misleading conclusions when comparing simulations with observations. 
In this paper, we have applied a new method to produce `observed' galaxies identified in mock imaging of hydrodynamical simulations. 
We {generate} low-redshift mock galaxies {from the TNG100-1 simulation of IllustrisTNG} and {analyse} them using standard `observational' techniques to extract their main structural parameters.
{We show that our technique can produce realistic {surface-brightness} distributions of the simulated galaxies, including classical morphological substructures, {such as} spiral arms and bars.} 
In particular, we find a very good agreement of the total luminosity and stellar mass {versus} halo mass {relationships,} and the galaxy stellar mass {versus} size {relationship between} mock observations and real galaxies.
{ We also compare the luminosity function and the mass function of the mock galaxy sample with literature data and find a good agreement at all luminosity and mass scales.}
In particular, { we find no significant tension {at} the bright end of the galaxy luminosity function, as reported in many analyses using simplified recipes to identify galaxy haloes, {which in fact miscount} the contribution of the extended galaxy haloes around large galaxies.
This demonstrates the critical impact of using observational driven approaches to the simulation analyses to produce realistic predictions to compare to observations.}

\end{abstract}
%-------------------------------------------------------%

%-------------------------------------------------------%

\section{Introduction}
\label{introduction}
{Large-scale} {hydrodynamical simulations} provide a direct way to study the properties of galaxies over a scale encompassing several decades of magnitudes in luminosity, mass, physical size and cosmic time \citep[e.g.][]{Evrard2002, Springel2005a, Vogelsberger2014a, Schaye2015}.
They have reached a level of {detail high enough to allow} a direct comparison {between the} physical properties of the simulated and observed galaxies, and, ultimately, {to} fully test the underlying galaxy formation scenario. 
However, {it is difficult to} make this comparison homogeneous, as converting the information encoded into the simulated particles defining galaxies in simulations {to} observable analogs measured in real galaxies (e.g. total luminosities in different bands, size, age, metallicity, etc.) is a complex task \citep[e.g.][]{Bottrell2017a, Bottrell2017b, Genel2018, Rodriguez-Gomez2019, Remus2021, Kudritzki2021}.

In hydrodynamical simulations, stellar substructures are usually identified by the {subfinder} algorithms \citep{Springel2001, Dolag2009} and subsequently treated as `galaxies' after{ the removal of} unbound particles.
This latter approach, though, has been {shown} to produce `galaxies' including an {excessive} amount of diffuse light or intra-cluster light \citep[ICL, see e.g.][]{Tang2020, Contini2020}, { hence adding a non-negligible extended component to the {surface-brightness} profiles} of brightest cluster galaxies {(BCGs)}, in particular \citep[see e.g.][]{Pillepich2018b}. 
On the other hand, observed galaxies are identified and characterized through their luminosity profiles. {For example,} their sizes are defined based on specific {surface-brightness} values or using experimental models to fit the radial brightness profile of the galaxy \citep[e.g.][]{Gonzalez2000, Zibetti2005,  Yang2009, Burke2015}.  
These differences in the galaxy definitions are a primary factor driving the discrepancies between observational quantities derived from simulations and the {those} observed in real galaxies. 

There have been {various} attempts to solve these discrepancies, {for example} by extracting quantities from simulations using observational methods.
{For instance, to compare galaxy properties from {the} TNG100-1 simulation {of IllustrisTNG} with Sloan Digital Sky Survey (SDSS), \cite{Pillepich2018b} used an aperture cut of $30\kpc$, as a simple approximation of Petrosian magnitudes \citep{Strauss2002}, to define the simulated galaxies.}
{This choice was empirically motivated to reproduce SDSS--like pipeline for galaxy aperture photometry (see also \citealt{Schaye2015} for a discussion), allowing {the derivation of} a galaxy stellar mass function in agreement with observations.}
{However, using a fixed aperture is a rather simplistic option as this tends to capture different fractions of the total stellar mass for large massive galaxies and small less massive ones, or {to} produce strong contamination from companion galaxies, especially in dense and complicated environments, and it cannot be used to define objects at different redshifts, considering the evolution of galaxy size.}
{\cite{Nelson2018} accounted for the attenuation of stellar light by dust when studying the colour bimodality of galaxies by subfinder algorithm at low redshifts in the TNG100-1 simulation of IllustrisTNG. This allowed them to obtain a better agreement with observations \citep{Alam2015} than previously found in other Illustris simulation analyses (\citealt{Vogelsberger2014b}), although an obvious offset with respect to SDSS galaxies at $z<0.1$ seems to remain unsolved.
This can be possibly due to still unaccounted for negative radial gradients of BCG+ICL colours \citep[e.g.][]{DeMaio2018, Montes&Trujillo2018, Contini2019}.
\citet{Genel2018} studied galaxy size--stellar mass relationship. 
They found that the sizes of low- and high-mass galaxies are much larger than the sizes in observations, especially for red/quenched galaxies.}

{Another way to understand the discrepancies between observations and simulations is to implement mock observations to add observational realism in simulations.
For instance, \citet[][]{Bottrell2017a, Bottrell2017b} used mock observation from Illustris simulations to study galaxy morphology.
They obtained the bulge + disc decomposition and galaxy half-light radii. However, these galaxies are still identified as substructures defined by subfinder procedure, rather than being selected on the basis of observational criteria.
Moreover, they work on mass maps, rather than light maps, with consequent loss of `observational realism'. 
Galaxy  morphology was also investigated on mock Pan-STARRS observations based on IllustrisTNG simulations in \citet{Rodriguez-Gomez2019}. 
These authors explicitly used non-parametric morphological diagnostics and performed two-dimensional S\'ersic fitting of galaxies. 
They found overall that optical morphologies of IllustrisTNG galaxies are in good agreement with observations, but could not reproduced the observed morphology--colour and morphology--size relations.
More sophisticated mock observations based on the TNG100 run of the IllustrisTNG project are also presented in \citet{Merritt2020}, who aimed to explore potential solutions to the low accreted mass fraction of galaxy outskirts and found it difficult to reconcile the differences between the observed and simulated galaxy outskirts. 
These are clear improvements with respect previous mock observational works \citep[e.g.][]{Brook2011, Agertz2011, Furlong2017}, in which there was no attempt to directly reproduce a mock imaging of simulated galaxies that would allow the reproduction of realistic surface-brightness profiles of stellar light.}

{In this paper, we want to test how improved observational realism can help the match of simulation predictions and observations. 
We focus, in particular, on the galaxy luminosity function (GLF) and stellar mass function (GSMF) to compare with results obtained from the Sloan Digital Sky Survey \citep{Yang2009, Bernardi2013, D'Souza2015}. 
We use the method developed in \citet{Tang2020}, optimized to reproduce the observational conditions of the data set one wants to compare with the observations.}
The novelty of our approach consists in producing two-dimensional (2D) projected images of the star-light in the simulation and then extracting galaxies following an approach similar to the observed source extraction in astronomical images.
Similarly to in observations, galaxies are identified and characterized only via their surface-brightness and integrated light properties, such as magnitudes, colours and sizes can be compared to equivalent quantities from observations.
{ Because we are interested in studying the statistics of galaxy populations (GLF and GSMF), we will first validate our mock galaxy sample against general galaxy properties such as galaxy colour distribution \citep[e.g.][]{Taylor2015}, galaxy star formation rate \citep[e.g.][]{Behroozi2013}, galaxy mass--dark matter halo mass relationship \citep[e.g.][]{Gonzalez2013, Ardila2021}. 
We also study the galaxy size--mass relationship \citep[e.g.][]{Shen2003} and compare our approach with observations \citep[e.g.][]{Baldry2012, Roy2018} and other studies also investigating this scaling relationships in simulations \citep[e.g.][]{Genel2018}.}

{The GLF and GSMF have been addressed in various state-of-the-art cosmological simulations (e.g. Illustris, \citet{Vogelsberger2014b}, Virgo Consortium's Evolution and Assembly of GaLaxies and their Environments (EAGLE) project, \citet{Schaye2015}, Horizon-AGN, \cite{Dubois2014}, IllustrisTNG,\citet{Pillepich2018b}, and similar projects \citep[e.g.][]{Springel2005b,Hopkins2014,Skillman2014,Cui2018}).}
{In particular, \citet{Vogelsberger2014a} investigated the GLF and GSMF in Illustris simulations and found a rough agreement with observations at the bright and massive ends.}
They stated that the active galactic nucleus (AGN) feedback is insufficient, and the volume of simulation box is too small, {causing an excess of galaxies at massive end.}
Using the next generation of Illustris simulations, i.e. IllustrisTNG,  \citet{Pillepich2018b} studied the stellar mass content of galaxy groups and clusters.
{they showed that the GSMF is  closer to the observational results at the massive end \citep[e.g.][]{Baldry2008, Baldry2012, Bernardi2013, D'Souza2015} than in previous Illustris-1 simulations.}
{It was also found that GSMFs in both simulations within a fixed aperture of $30 \kpc$ are obviously lower than those within the twice stellar half-mass radius at the massive end at redshift z = 0.}
The aforementioned results illustrate that a reasonable galaxy definition is important in the study of galaxy populations, especially if the aim is a comparison with observational results. 
In this paper, we argue that the comparison of simulation results with observations should be performed with caution and can be trusted only if the same, or at least very similar, way to define galaxies has been used.

This paper is organized as follows. 
{In Section \ref{simulation} and \ref{methods}, we introduce the simulation and our galaxy definition.
In Section \ref{population}, we show the galaxy colour and star-formation rate distribution, galaxy stellar mass/luminosity--halo mass relation, and galaxy stellar mass--size relation.
We represent our results and the comparison, focusing on the GLF and GSMF, in Section \ref{GLF_GSMF}.
We detail some systematics in Section \ref{systematics}.
Conclusion is given in Section \ref{conclusion}.}
%-------------------------------------------------------%

\section{simulations}\label{simulation}
In this paper we will use the {TNG100-1 simulation of IllustrisTNG} \footnote{https://www.tng-project.org}.
This is a cosmological hydrodynamical simulation with $1820^3$ dark matter and $1820^3$ gas particles, in a cubic box of $(110.7\Mpc)^3$.
The mass resolution of the simulation is $1.4\times10^6\Msun$ for baryon particles, while the spatial resolution is set by the Plummer softening length, which is $0.74\kpc$. 
The cosmological parameters are taken from WMAP-9 \citep{Hinshaw2013}: $\Omega_m=0.2726$, $\Omega_{\Lambda}=0.7274$,  $\Omega_{b}=0.0456$,  $n_s=0.963$, $\sigma_8=0.809$, and $h=0.704$.
{The detailed description of the data base we utilized can be found in \cite{Nelson2019}. We also refer the interested reader to 
the paper series on TNG100/300 for a more detailed description of the stellar content \citep{Pillepich2018b}, galaxy clustering \citep{Springel2018}, galaxy colours \citep{Nelson2018}, chemical enrichment \citep{Naiman2018}, and magnetic fields \citep{Marinacci2018}.}
Compared to previous Illustris simulations \citep[][]{Vogelsberger2014a}, {the TNG100-1 simulation of IllustrisTNG} includes a revised AGN feedback model, controlling the star formation efficiency of massive galaxies \citep{Weinberger2017}, and galactic wind model, whose feedback inhibits the star formation efficiently in low and intermediate mass galaxies \citep{Pillepich2018a}.

\section{methodology}\label{methods}
We use the three-dimensional (3D) spatial datacube of the {TNG100-1 simulation} to build the catalog of galaxies in the simulation volume. 
The procedure is based on the conversion of {the 2D projection of the simulated particles} into a pixellized image, where each individual pixel contains the information of the integrated light carried by all the star particles in the volume behind it. 
Below we detail the various steps of the procedure.

\begin{enumerate}
\item Projection of groups. 
The first step is the projection of the stellar particles of the dark matter halo grouped by the Friend-of-Friend (FoF) algorithm \citep{Davis1985} {to produce the stellar surface-brightness images} on three projecting planes, $xy,\ yz,\ zx$.
Note that the use of dark matter halo as projecting group allows us to minimize the projection effect.
We divide each FoF group into 2D grids on each projecting plane, where each grid cell corresponds to a {\small CCD} pixel. 
At a given redshift $z$, the grid size $D$ and the angular-diameter distance $d_{A}$ are related trough the relation
\begin{equation}
	\label{eq_D}
	d_A=\frac{ D }{ \alpha  }=\frac{ a_0 r }{ 1 + z },
\end{equation}
while the luminosity distance $d_{L}$ is given by
\begin{equation}
	\label{eq_L}
	d_L= a_0 r ( 1 + z ),
\end{equation}
where $\alpha, \ a_{0}, \ r$ are {\small CCD} pixel scale, scale factor at the present time and proper distance, respectively. 
Using the Friedmann Equation, $a_{0}r$ is defined by
\begin{equation}
	\label{eq_ar}
	a_{0} r=\frac{ c }{ H_{ 0 } }\int_{ 0 }^{ z } { \frac{ dz }
	{[ \Omega_{ \Lambda,0 } + \Omega_{ m,0 }( 1+z )^3 ]^{1 / 2} } },
\end{equation}
where $H_{ 0 }$, $\Omega_{ \Lambda,0 }$ and $\Omega_{m,0}$ are the present time Hubble constant, and the dark energy and total mass  cosmological parameters, respectively.
With the estimated grid size $D$, the star particles inside the group radius are binned into a 2D mesh on each projecting plane.
{Note that the star particles in each grid are simply summed by a histogram approach.
Thus, the position of the grid is used instead of the mean position of all the star particles in this grid.}
The properties of each grid are then weighted by either luminosities or stellar masses.
The surface brightness in any band $x$ for each pixel is given by
\begin{equation}
	\label{eq_ux}
	\mu_x=-2.5\log\frac{I_x}{L_{\odot,x} \cdot pc^{-2}}+21.572+M_{\odot,x},
\end{equation}
where $L_{\odot,x}$ and $\rm M_{\odot,x}$ are the solar luminosity and the absolute solar magnitude in the $x-$band and $I_{x}$ is defined as
\begin{equation}
	\label{eq_lx}
	I_x=\frac{L_x}{\pi^2 D^2}(1+z)^{-4},
\end{equation}
where $L_x$ is the luminosity in a given pixel of size $D$ in the same $x-$band. 
The {TNG100-1} simulation of IllustrisTNG provides all the stellar population information related to the individual mesh, and hence we can collect the following astrophysical quantities: 
(1) the total stellar mass and luminosity as a sum of the masses and luminosities (U, B, K, V, SDSS-g, SDSS-r, SDSS-i, SDSS-z bands) of all star particles in the mesh;
(2) the mean age and metallicity calculated by weighting the stellar luminosity;
(3) the sum of the individual star-formation rates of all gas cells in the mesh.

%---------------figure_illustration-------------------------%
\begin{figure*}
\centering
\includegraphics[width = 0.99\textwidth, height= 15cm]
{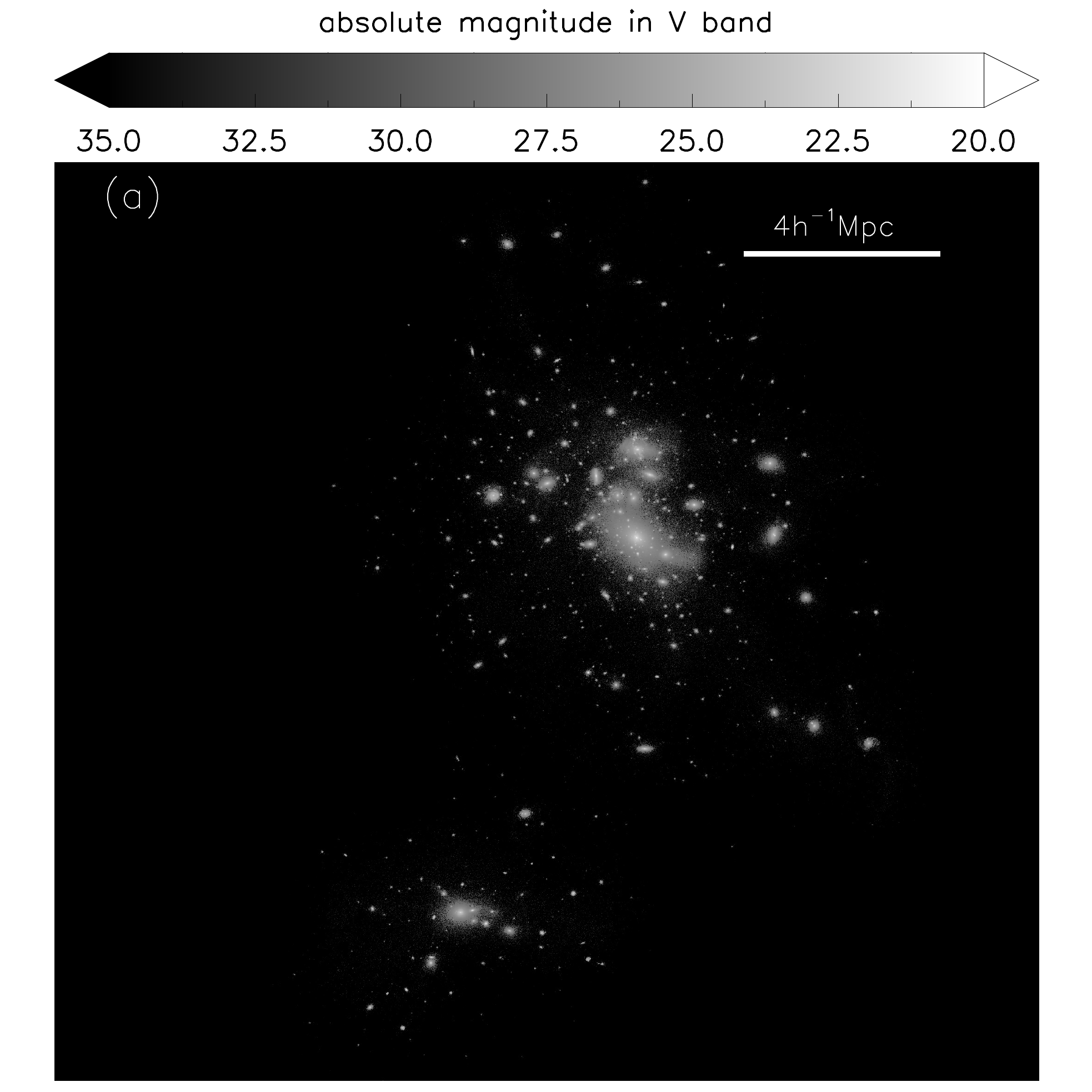}
\includegraphics[width = 5.85cm, height= 5.85cm]
{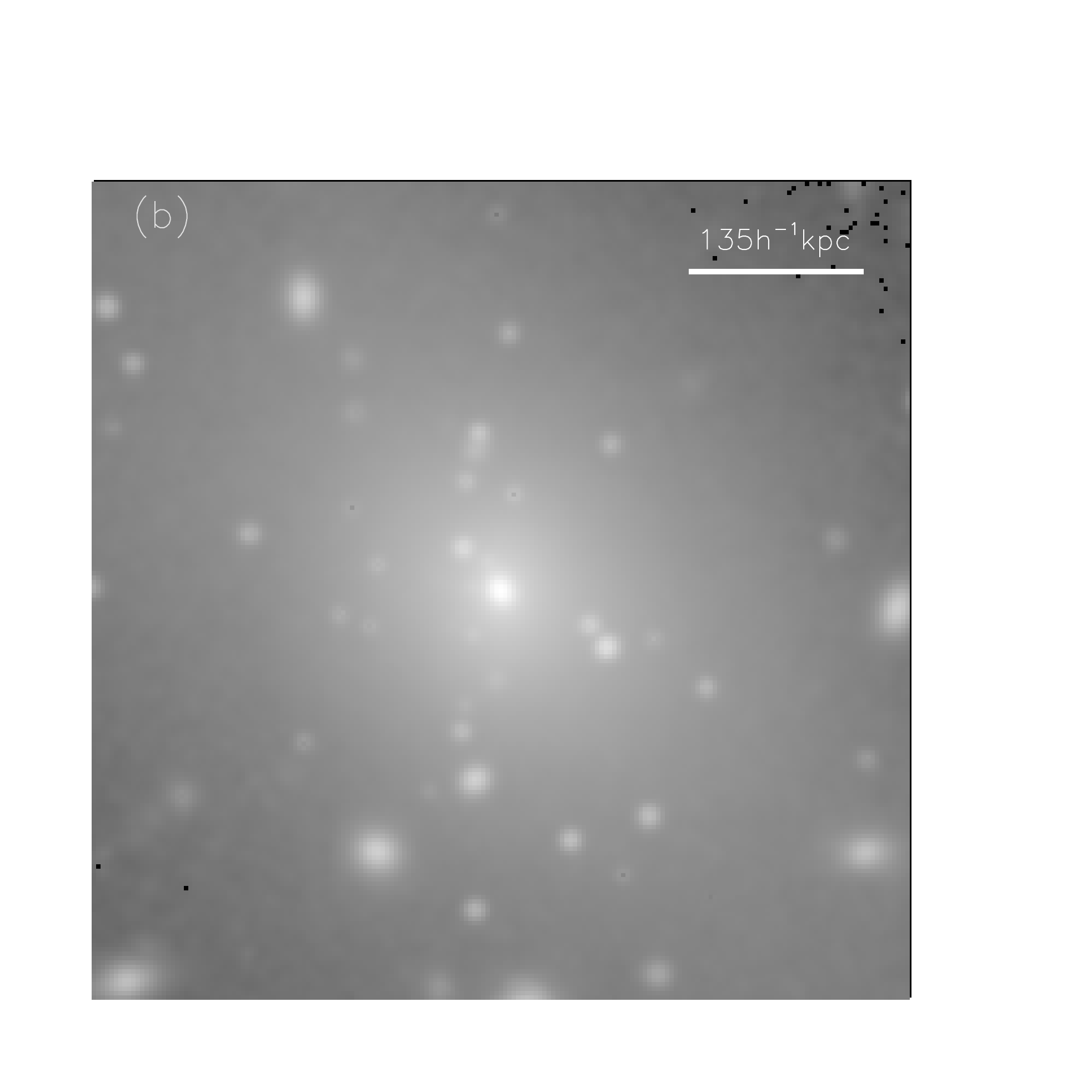}
\includegraphics[width = 5.85cm, height= 5.85cm]
{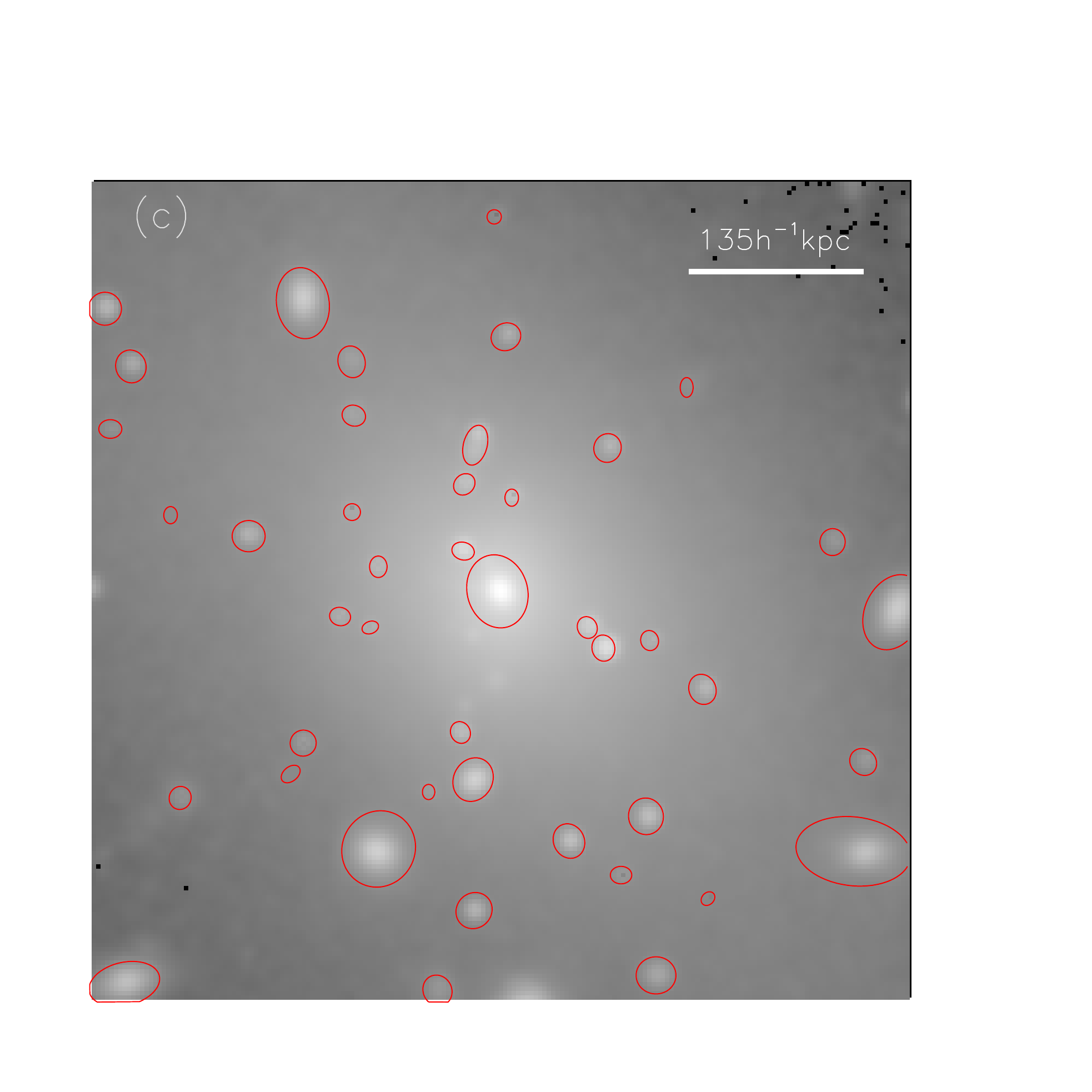}
\includegraphics[width = 5.85cm, height= 5.85cm]
{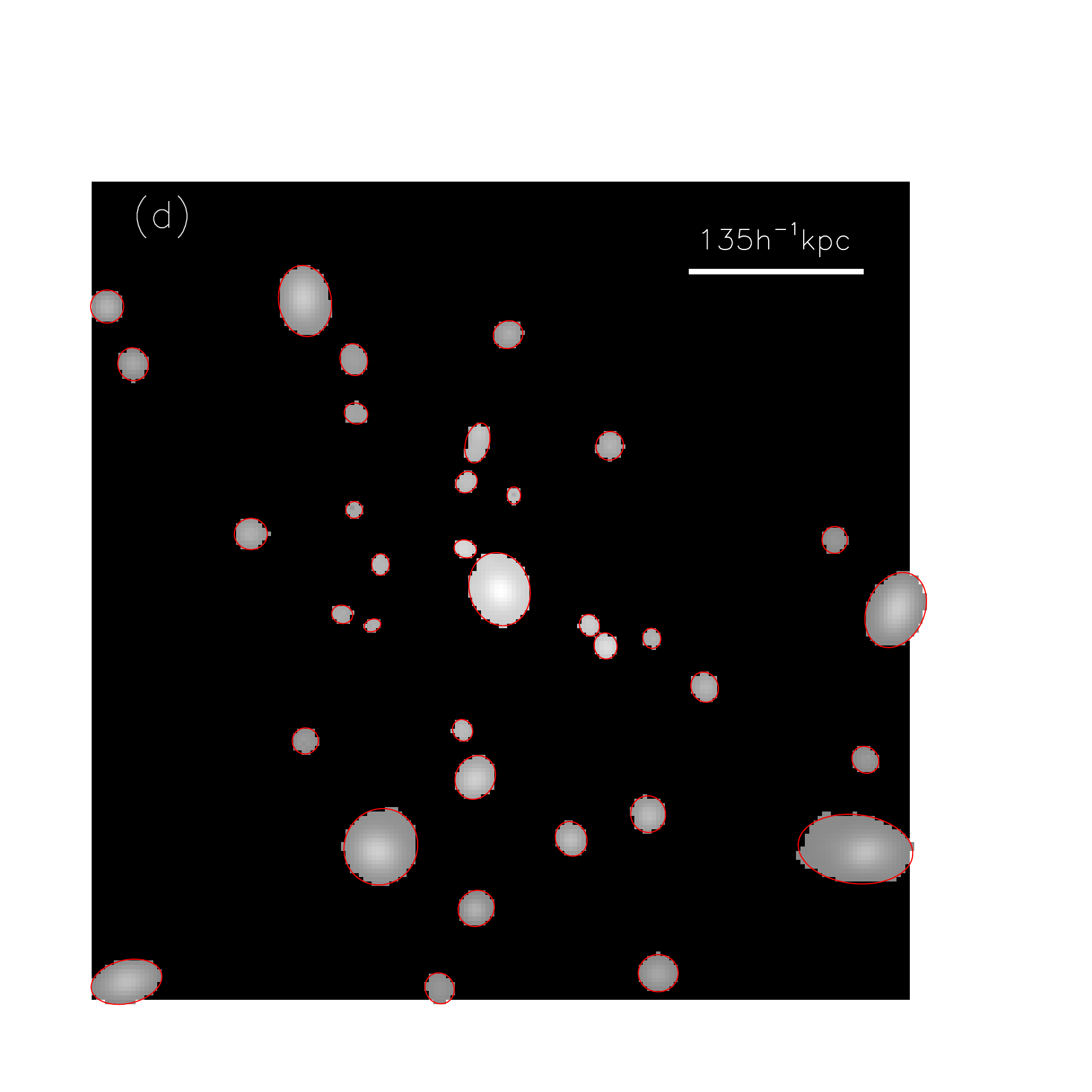}
\hspace*{-20pt}
\caption{A simple illustration of our new galaxy definition convolved with the simulation-PSF. 
{All the figures are light images in the V band.}
a): the projected image on x-y plane of the most massive group in snapshot at $z=0 \ (move \ to \ 0.01)$, with size of $4\Mpch$ by side; b): the central region of $135\kpch\times135\kpch$; c) the galaxy candidates extracted by the surface-brightness peak and large enough galaxy edge defined by reconstruction of a series of surface-brightness criterion; d) the galaxies with stellar mass greater that $10^{8}\Msun$ in central region, masking the diffuse stellar component and background. 
}
\label{fig_illustration}
\end{figure*}
%------------------------------------------------------%
%---------------figure_galaxy-----------------------------%
\begin{figure*}
\centering
\includegraphics[width = 0.49\textwidth]{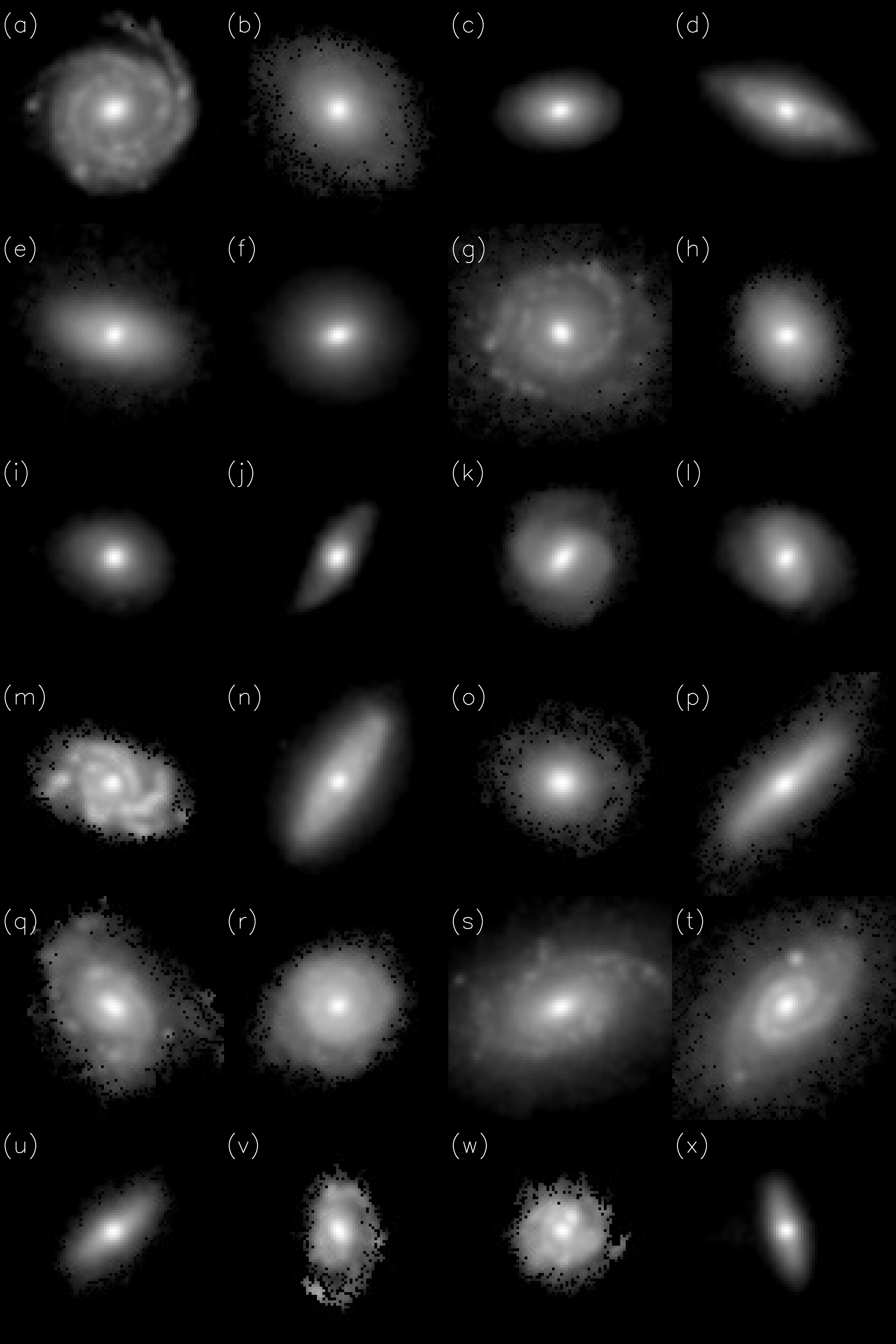}
\includegraphics[width = 0.49\textwidth]{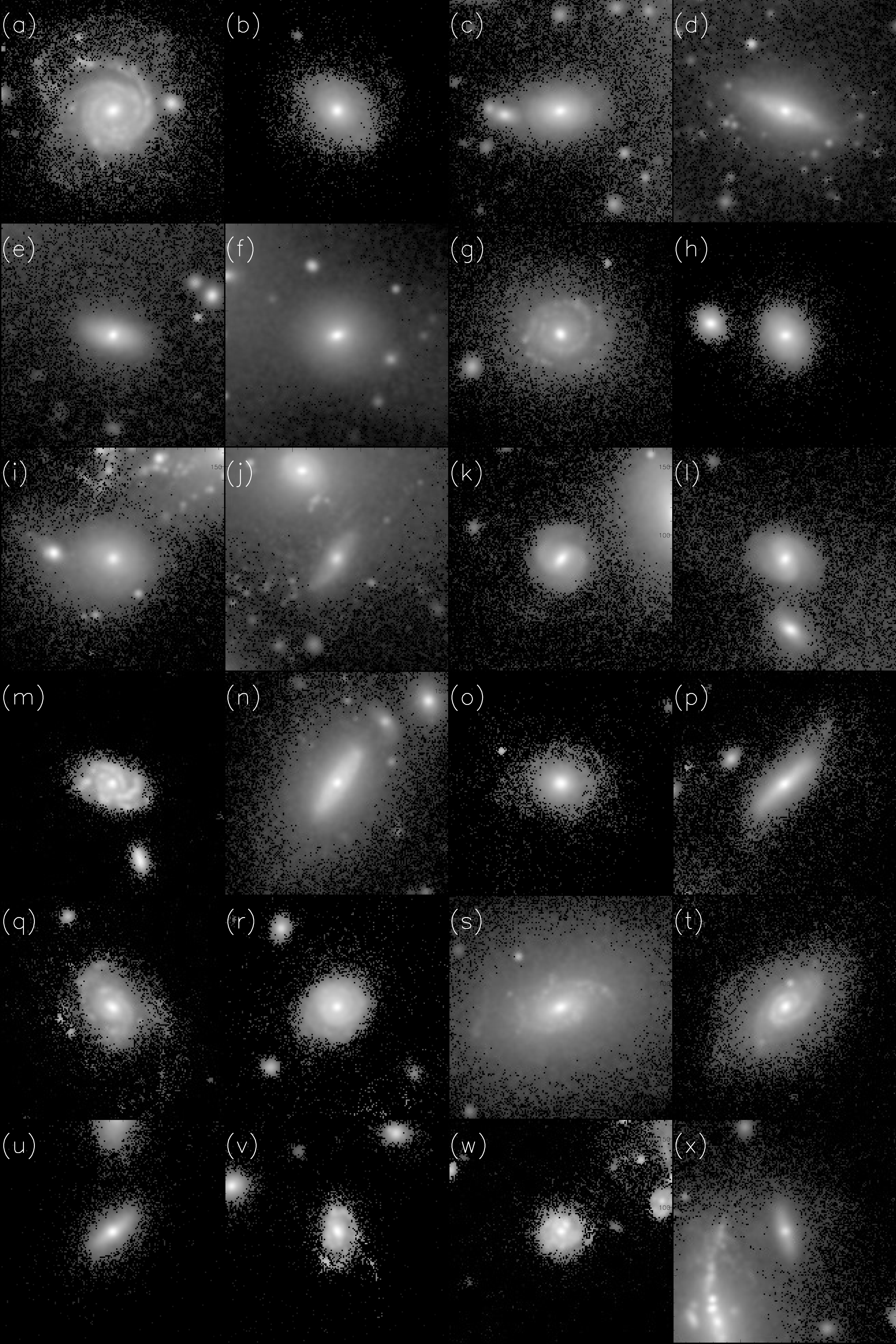}

\caption{{\it Left-hand panel}: luminosity profile in V band of 24 galaxies extracted from the galaxy sample, with projected region of $60\kpch \times 60\kpch$ for each galaxy; {\it Right-hand panel}: luminosity profile of the all stellar component, with projected region of $120\kpch \times 120\kpch$ for each galaxy.
All the illustrations are centered at each galaxy luminosity peak and extracted from the snapshot at $mock\ z=0.01$.
All galaxies have stellar mass larger than $10^{10}\Msunh$.
}
\label{fig_galaxy}
\end{figure*}
%-------------------------------------------------------%

\item Convolution with point-spread function (PSF).
The PSF width ($\omega$) and CCD pixel size ($\alpha$) are two important parameters to account for when comparing observations with simulated galaxy properties.
In \citet{Tang2018}, we have demonstrated that the CCD pixel size has a minor impact on galaxy measurements in simulations, while the PSF width has a major impact on the derived quantities, especially for faint systems/features.

As anticipated in Section \ref{introduction}, in this work we want to compare our results with SDSS, as a standard reference data set, especially for lower-redshift galaxy statistics and scaling relations. 
In terms of pixel scale, the SDSS CCD pixel size (0.396$''$) is less than 1/10 of the size of the gravitational softening length (0.74 kpc) at $z=0$, while it becomes comparable to this latter at higher redshift.
For instance, the softening length would correspond to a PSF of 5$''$, 0.5$''$, 0.3$''$ for simulation snapshots at $z=0, \ 0.1,\ 0.2$, respectively.
Hence, setting the fiducial simulation-CCD pixel size of the projected images as the one corresponding to the softening length is a good compromise to match the observation pixel scale at redshifts larger than $z=0$, while it might provide smoother images at $z=0$ than really observed in SDSS images.
However, we just remark that the $z=0$ case is ideal, from the observational point of view, because this would correspond to a `local' universe, where the PSF has a minor impact on the surface-brightness observations of nearby galaxies.
For all cosmological results (at $z>0$) our PSF choice is rather realistic. 
However, in order to check the impact of the `simulation-PSF' obtained above by converting the softening length into an angular resolution, we also consider a so-called `SDSS-PSF', whose width is given by the average SDSS fullwidth half maximum (FWHM$=\omega=1.43''$).

As PSF model (either using the simulation-PSF or the SDSS-PSF), we adopt the Moffat formula \citep{Moffat1969, Trujillo2001a}, which is a function of FWHM (F) and $\beta$:
\begin{equation}
	\label{MOFFAT_PSF}
f(x,y)=4(2^{1/\beta}-1)\frac{\beta-1}{\pi F^2}[1+4(2^{1/\beta}-1)\frac{x^2+y^2}{F^2}]^{-\beta},
\end{equation}
where $\beta=4.765$ and $F^2=8(\omega/\alpha)^2$. 

{At this stage, we also add a simulated sky background and CCD noise using the IDL routine {\it PoiDev} to obtain a Poissonian realization of both maps \citep[e.g.][]{Civano2016}.
For each CCD pixel, the variance of the Poissonian distribution is set to $30$ in units of $\Mag$ and magnitude of CCD pixel, for the sky background and CCD noise, respectively.}

\item Segmentation of the galaxy sample. 
The procedure to identify and separate the individual galaxies in the 2D image of the simulation {obtained by above steps (i) and (ii) }follows the prescription from \citet{Tang2020}, which we summarize here. 

(1) {First, we identify the galaxies} from the mock observation image by applying a series of surface-brightness limits (SBLs), namely $18$ to $26.5 \Mag$ with surface-brightness intervals of $0.1 \Mag$.
These upper and lower bounds are chosen because $26.5 \Mag$ level is the limiting surface brightness of SDSS corresponding to $\sim$1\% of the sky brightness, while $18 \Mag$ is the maximum surface brightness measured in all our virtual CCD pixels.
(2) Then, we obtain the catalog of sources for each SBL sample and compare, in turn, the galaxies defined by two neighboring SBLs (the upper SBL, 0.1$\Mag$ btighter, and the lower SBL, 0.1$\Mag$ fainter) and build a temporary catalog of mock galaxies that are simultaneously defined by the two nearby SBLs.
Each `galaxy', defined as a close contour in a given SBL, is kept in the catalog related to that particular level, if it is still defined as a single galaxy in the upper (brighter) contour. 
If on the other hand a galaxy splits in two or more systems at the brighter SBL, then it is excluded from the temporary catalog of the lower SBL and the subsystems found in the upper level are included in the updated catalog of galaxies.
(3) Finally, repeating (1) and (2) steps for each brighter SBLs, we obtain the final sample of mock galaxies (see also Figures 2 and 3 in \citet{Tang2020} for a summary of this segmentation procedure).

{In Fig.~\ref{fig_illustration} we show the result of this SBL segmentation procedure (SBLSP, hereafter) on the most massive group in the simulation}, as an illustrative example. 
In the bottom row, in particular, we see the details of the galaxies selected according to the lowest SBL defined to match with the SDSS limits. 
There are a few low-surface-brightness objects that remain undetected by the procedure, just because they are below the adopted threshold ($26.5 \Mag$).
{In Section \ref{test}, we will use a standard pipeline for source extraction in astronomical data, for example SEXTRACTOR \citep{Bertin&Arnouts1996}, to verify the reliability of the SBLSP.}

\item Dust extinction. 
This is a crucial component to account for when characterizing galaxy luminosities.
This is true also when reproducing realistic mock observations. 
There have been many attempts to emulate dust extinction in simulations, using various dust models \citep[e.g.][]{Nelson2018, Rodriguez-Gomez2019, Vogelsberger2020}.
In this work, we use the radiative transfer code SKIRT \citep[][]{Baes2011} to add a dust component to the star-forming mock galaxies identified in the procedures (i)-(iii). 
First, we determine the morphological parameters, i.e. axes and inclination of our star-forming sample.
{Then, using the monochromatic simulation of a dusty disc galaxy with SKIRT \citep{Yuan2021}, with the assumption of the  Calzetti attenuation curve \citep{Calzetti2000}, we produce the expected dust contribution for face-on or edge-on projections.
Finally, according to the morphological parameters, we associate the dust distribution produced by SKIRT to each mock galaxy.
The luminosity of each CCD pixel of star-forming mock galaxies is equal to the original luminosity minus the dust extinction in the same place as the mock shape.}
\end{enumerate}

In Fig.~\ref{fig_galaxy}, we illustrate a gallery of mock galaxies with masses greater than $10^{10}\Msunh$, selected by our algorithm in the virtual images of the simulation. 
We can clearly see details of their different morphology (left-hand panels), ranging from featureless spheroids (early-type galaxies) to spiral arms and star-formation regions in discs (late-type galaxies). 
In the same Figure we show more details of their environment in the zoomed-out images (right-hand panels).
In particular, we distinguish close substructures that do not belong to the main galaxy in, for example, panels (a), (c), (e), (f), and (x).
For dense regions, our method can naturally distinguish the diffuse stellar from galaxies, for example in panels (e), (g), (n), (s), and (t).
We also find enormous envelopes for some massive isolated galaxies, which are generally removed in our galaxy samples.
In Section \ref{radius}, we will discuss the galaxy half-light radii--stellar mass relationship.
In a future work, we will address specifically galaxy morphology and surface photometry using mock galaxy imaging in more detail, which is beyond the purpose of this paper.

%---------------figure_color_sfr----------------------------%
\begin{figure*}
\centering
\includegraphics[width = 0.49\textwidth]{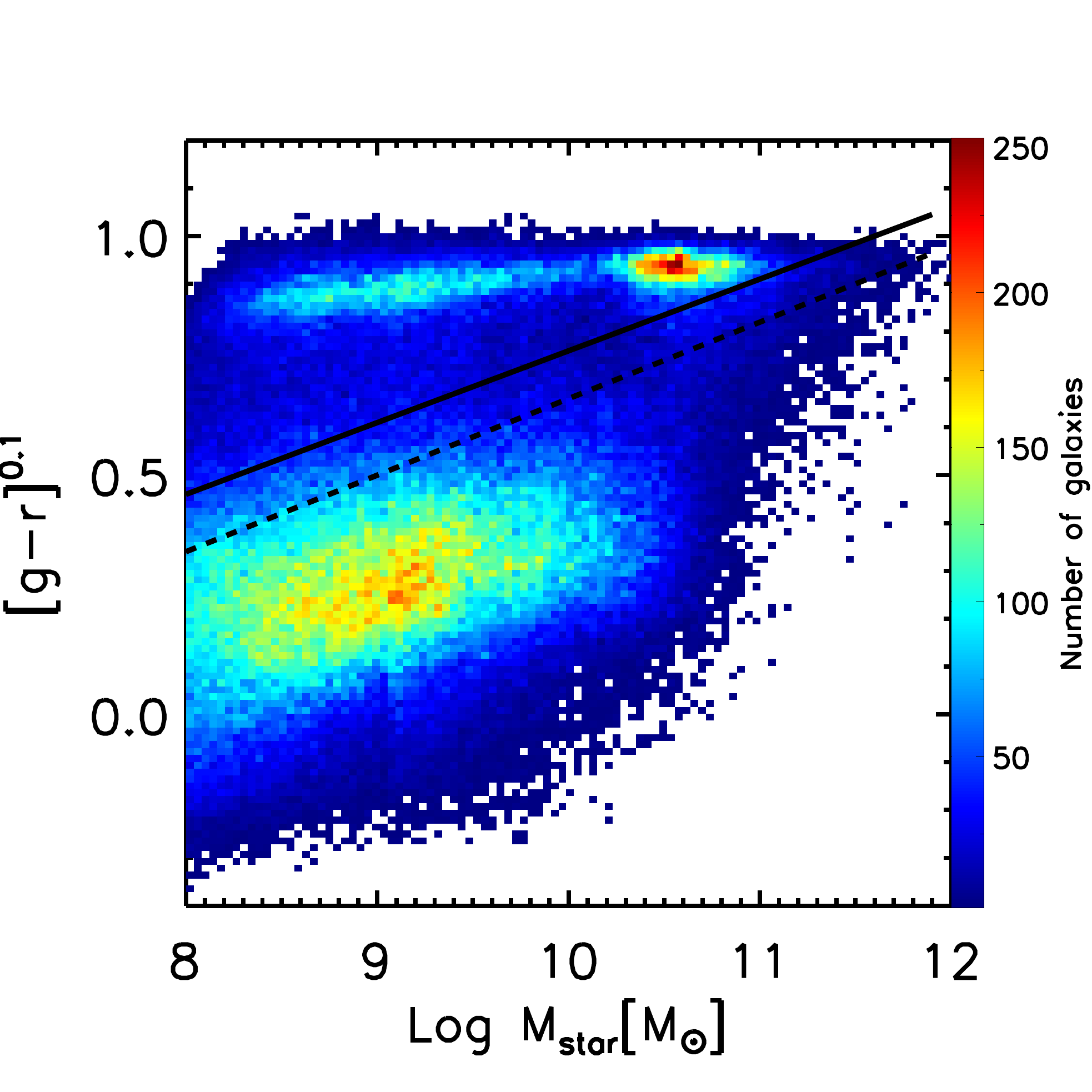}
\includegraphics[width = 0.49\textwidth]{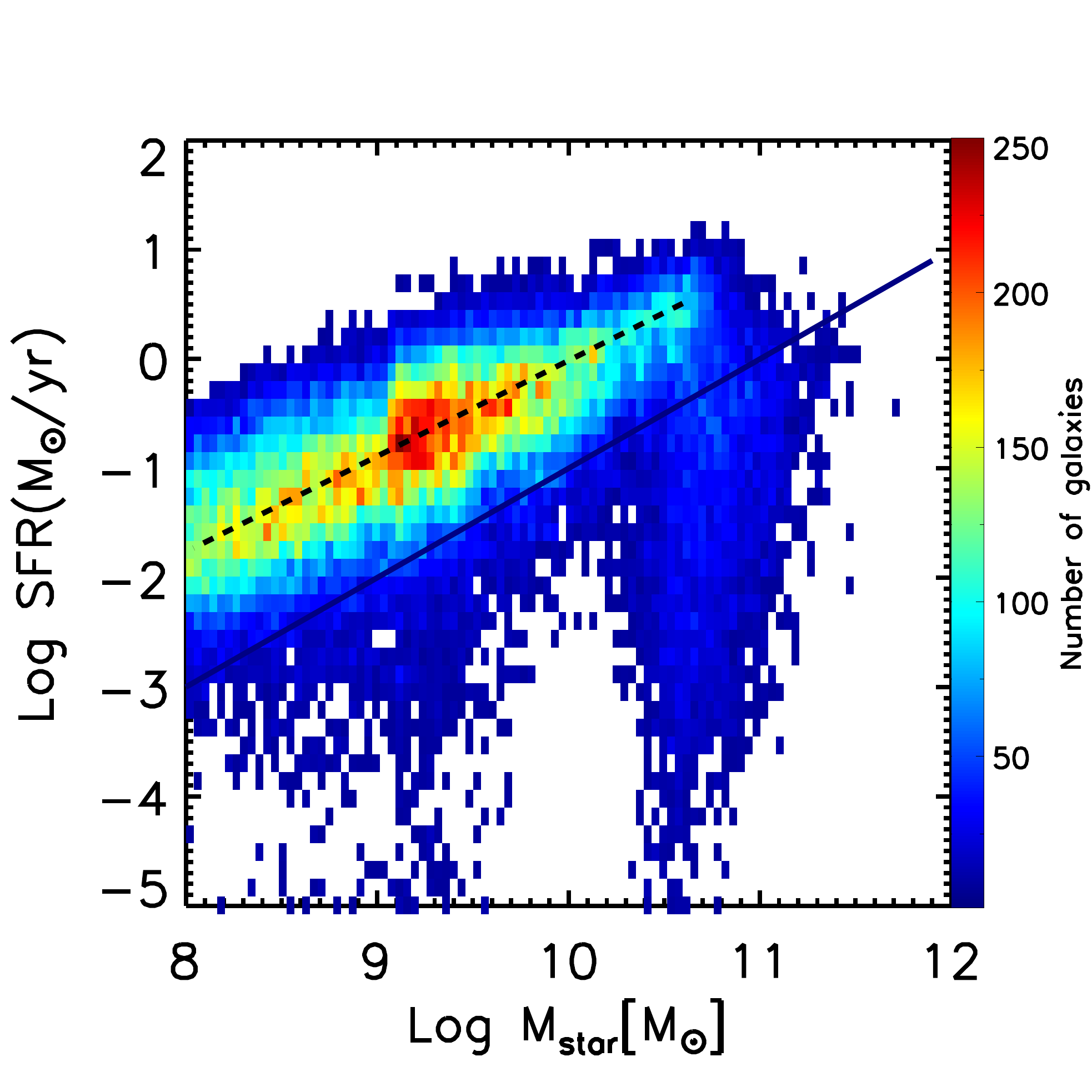}
\caption{
{\it Left-hand panel}: { $[g-r]^{0.1}$ colour--mass diagram of mock simulated galaxies. 
The lines correspond to the linear fit formula, $[g-r]^{0.1}=\alpha+\beta (\log {\rm M}_{\ast}-10)$.
The dashed line is the best fit to our galaxy sample (with best-fitting parameters $\alpha=0.56$, $\beta=0.16$), and the solid line is the results for the SDSS galaxies from \citet[][]{vandenBosch2008} (with best-fitting parameters $\alpha=0.76$, $\beta=0.15$).}
{{\it Right-hand panel}: SFR--stellar mass diagram of mock simulated galaxy.
The black line represents ${\rm SFR/M}_{\ast}=10^{-2} Gyr^{-1}$ to define star-forming ($>10^{-2} Gyr^{-1}$) or quenched galaxy ($<10^{-2} Gyr^{-1}$), which is adopted in EAGLE \citep[][]{Schaye2015},  IllustrisTNG \citep[][]{Donnari2019}, and other literatures.}
{The dashed line represents the star-formation main sequence, which is obtained by the maximum distribution of SFR within stellar mass bins.
The dashed line has a slope equal to $\sim 0.89$, meaning that $\log{SFR} \propto 0.89\log{\rm M_{\ast}}$.}
}
\label{fig_color_sfr}
\end{figure*}
%-------------------------------------------------------%

\section{Galaxy properties}\label{population}

When comparing the statistics of galaxies from different data sets (e.g., their luminosity or mass function) or galaxy scaling relationships (e.g., the size--luminosity, or size--mass, or the so called Kormendy relationship, namely the the surface-brightness--size relationship), one needs to take  into account carefully the observational limits of each data set, which can be quantified by their limiting luminosity, limiting mass or limiting surface brightness. 
This is also true in comparisons of observations with simulations, as incomplete samples can affect the behaviour of the galaxy counts (by definition) or the trend of the scaling relationships. 
It goes without saying that observational limits (for, absolute luminosity, stellar mass, for example), depending on the redshift, have to be derived for the mock observations at each simulation snapshot.

In this work, because we want to compare the mock observations of {IllustrisTNG galaxies} with observations from SDSS, we select a galaxy sample matching the same observational limit of surface brightness, namely 1\% of sky brightness, and magnitude, namely $17.77$ in the $r$ band, as SDSS \citep{Strauss2002}.
In addition, we produce mock images using simulation snapshots at three redshifts, $z=0$ (which, for it to be translated into angular sizes compatible with a telescope field-of-view and match the SDSS resolution--see Section \ref{methods}(ii)--is artificially moved \ to \ 0.01), 0.1, 0.2, to cover the same redshift range of SDSS.
As the snapshot at $z=0$ is shifted to $z=0.01$, {we assume that there will be no significant structure and galaxy evolution in this time range.}
{ In terms of total mass, we only consider galaxies with stellar masses above $1.4\times 10^8\Msun$, which match well the typical masses of the observed GSMF we want to compare to.
Moreover, the host dark matter halos are more massive than $10^{12}\Msun$.}

Before we proceed to discuss the statistical distributions of the galaxy properties as derived by the mock observations, we want to check whether the derived galaxy `observables' are consistent with scaling relationships from real galaxies. 
In this section, we discuss the stellar mass-- colour ({$[g-r]^{0.1}$, i.e. the $g-r$ colour rescaled to match the values at redshift $z=0.1$}) and mass--star formation rate (SFR) relationships for the galaxy sample from simulation mock observations.
These relationships are important, for example to separate blue from red galaxies and star-forming from quenched galaxies.
{We calculate $[g-r]^{0.1}$ using the conversions provided by \citet[][see their Table 2]{Blanton&Roweis2007} to obtain the colour $[g-r]^{0.1}=0.7088-1.3197[(g - r) - 0.6102]$, where the $g$ and $r$ are magnitudes in SDSS filters.
As mentioned at the end of Section~\ref{methods}(i), we have also obtained the SFR ``map" in the mocking images.
The SFR of a galaxy is simply defined by the total SFR of CCD pixels (namely the sum of the individual SFRs of all gas cells) belong to the galaxy.}
Once these two classes of galaxies are defined, we will investigate some basic scaling relationships such as the galaxy luminosity/stellar mass versus halo mass and the galaxy size versus stellar mass relationships and compare them against results from both observations and simulations applying the standard analysis tools based on subfinder grouping.

\subsection{Galaxy colour and star formation rate}\label{color_sfr}

Galaxy colours have been {discussed for IllustrisTNG \citep{Nelson2018}, EAGLE \citep{Trayford2015}, among others}. 
Most of these works have reproduced the well-known galaxy colour bimodality, which is a distinctive observational features of galaxy evolution  \citep[e.g.][]{Baldry2004, Taylor2015} that is now reasonably well understood is simulations.

In the left-hand panel of Fig.~\ref{fig_color_sfr}, we plot the correlationship between the galaxy stellar mass and $[g-r]^{0.1}$ colour derived from our mock observations.
Following \citet{vandenBosch2008}, we divide red from blue galaxies using the simple linear relationship in the colour--mass diagram, i.e. $[g-r]^{0.1}=\alpha+\beta (\log {\rm M}_{\ast}-10)$. 
For our mock observations we obtain $\alpha=0.56$ and $\beta=0.16$, corresponding to the black dashed line in Fig.~\ref{fig_color_sfr}. 
In the same figure we also show the same relationship obtained for SDSS observation (\citealt{vandenBosch2008}, black solid line), where $\alpha=0.76$, $\beta=0.15$. 
The two relationships have the same slope, and despite a small shift in our relationship, due to a sharp cut at $[g-r]^{0.1}\sim 1$.
The two relationships are consistent with each other.
Furthermore, these relationships nicely split the bimodal peaks of the colour distribution of our galaxy samples.
There are few galaxies (< 8\% of the total number of galaxies) between those two lines, which only slightly affect the final statistics of galaxy colors. 
{In our work, our fitting line (i.e. the black solid line in the left panel of Fig.~\ref{fig_color_sfr}) is adopted to divide galaxies into red galaxy ($[g-r]^{0.1}>0.56+0.16 (\log {\rm M}_{\rm star}-10)$), and blue galaxy ($[g-r]^{0.1}<0.56+0.16 (\log {\rm M}_{\rm star}-10)$).}

In the right-hand panel of Fig.~\ref{fig_color_sfr}, we plot the correlationship between galaxy SFR and galaxy stellar mass, corresponding to the galaxy star formation main sequence.
{The main sequence (dashed line in the right-hand panel of Fig.~\ref{fig_color_sfr}) has a slope of 0.89, which is similar to that discussed for IllustrisTNG \citep{Donnari2019} and observations \citep[e.g.][]{Elbaz2007, Salim2007, Oliver2010, Zahid2012, Behroozi2013, Whitaker2014}.}
In Fig.~\ref{fig_color_sfr}, we separate star-forming from quenched galaxies using the specific star formation rate (sSFR) \citep[e.g.][]{Schaye2015,Donnari2019}.
The black solid line shows the value adopted, namely $\rm{sSFR=SFR}/{\rm M}_{\ast}=10^{-2} Gyr$, to make this split in the two samples, such that the star-forming galaxies have a value of sSFR above the threshold and the quenched galaxies have a value below.

\subsection{Galaxy stellar mass - size relation}\label{radius}
Galaxy sizes reflect the physical processes involved in their formation and evolutionary history.
For instance, galaxy sizes are known to be related to the angular momentum \citep[e.g.][]{Mo1998, Somerville2018}. 
Galaxy merging can destroy the galaxy morphology and affect the galaxy size \citep[][]{Robertson2006, Bezanson2009, Barro2017}. 
Gravitational instabilities inside galaxies may also lead to a change in the galaxy dimensions \citep[][]{Dekel&Burkert2014}.
In general, observations show that more massive galaxies have larger sizes than less massive systems, and late-type/star-forming galaxies tend to be bigger than early-type/passive ones of the same mass \citep[e.g.][]{Shen2003, Baldry2012, Roy2018, Miller2019}. 

The sizes of our mock galaxies are equivalent to the classical definition of the half-light radius, $R_h$, from `growth curve', namely
\begin{equation}
R_h=\frac{L_T(R_{\rm out})}{2},
\label{eq:r_h_growth}
\end{equation}
where $L_T$ is the total light within the radius, $R_{\rm out}$, defined as `boundary' of the galaxy (i.e. corresponding to 26.5 mag/arcsec$^2$ in our case). 
According to this definition, $R_h$ if the circular radius enclosing half of $L_T$, which corresponds to the circularized radius definition $R_m\sqrt{ab}$, if $R_m$ is the major-axis estimate from elliptical outermost boundary (as in Fig.~\ref{fig_test}). 
This is an important distinction, as in observations sizes are defined either as circularized or major axis (see \citealt{Roy2018} for a discussion on the impact of the different definitions).
Furthermore, in Equation \ref{eq:r_h_growth}, we cannot use the same boundary defined in our SBLSP discussed in Section \ref{methods}. 
Indeed, this is determined in the PSF-convolved images, while sizes in observational works are obtained using PSF-convolved models; that is, {the aim is to measure the intrinsic half-light radii of galaxies before they are `blurred' by the PSF (see, e.g. \citealt{Baldry2012}, \citealt{Roy2018}).}
 Hence, to reproduce the observational measurements, we apply the Equation \ref{eq:r_h_growth} to the simulated galaxies before the convolution with the PSF. 
%---------------figure_R---------------------------------%
\begin{figure*}
\centering
\includegraphics[width = 0.99\textwidth,height=8.5cm]{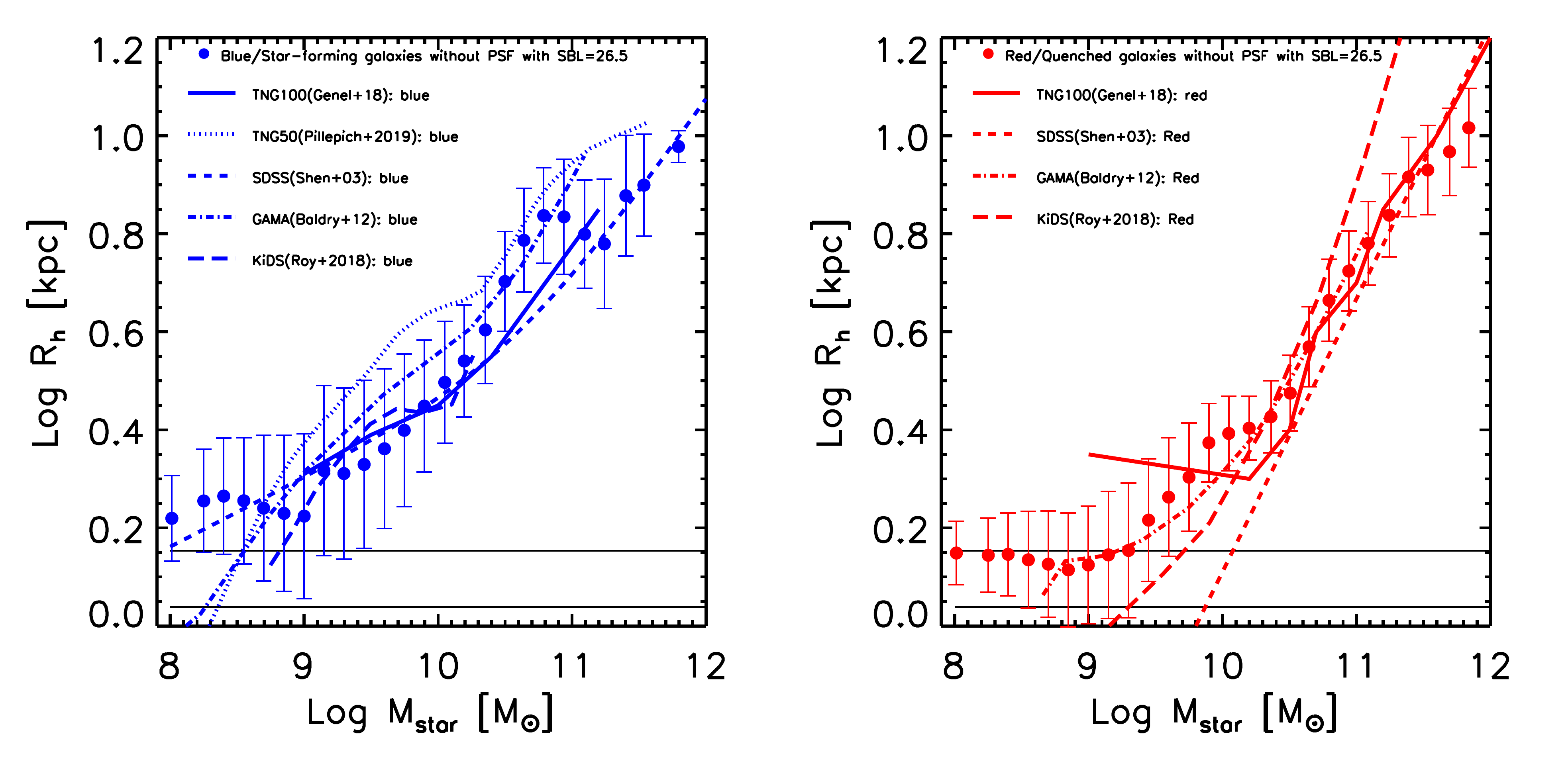}
\caption{
The galaxy stellar mass (${\rm M}_{\rm star}$)--size (${\rm R}_{\rm  half-light}$) relationship.
The colours of the lines denote the types of galaxies, blue for blue/star-forming galaxies, red for red/quenched galaxies.
The scaling relationship in \citet{Genel2018} from {TNG100-1} simulations at $z=0.1$ is shown as a solid line, {while the scaling relationship of star-forming galaxies from the TNG50 simulation \citep{Pillepich2019} is plotted as a dotted line in the left-hand panel.}
We also show results from observations in both panels from \citet[][short dashed lines]{Shen2003}, \citet[][dotted dashed lines]{Baldry2012} and \citet[][long dashed]{Roy2018}. 
{Note that \citet{Shen2003} separated red spheroids from blue disc galaxies using a S\'ersic index $n=2.5$.
The two horizontal lines denote the TNG100-1 softening length $R=0.74\kpch$, and $4\times0.74\kpch$, defining the spatial resolution of the simulation.}
}
\label{fig_R}
\end{figure*}
%-------------------------------------------------------%

In Fig.~\ref{fig_R}, we show the galaxy half-light radius--stellar mass relationship for blue/star-forming and red/quenched galaxies as defined in Section \ref{color_sfr} (see also Fig.~\ref{fig_color_sfr}). 
In the same figure, we also plot the relationship from observations (SDSS  (\citealt{Shen2003});  Galaxy And Mass Assembly (GAMA) survey, (\citealt{Baldry2012}); Kilo Degree Survey (KiDS) (\citealt{Roy2018})) and previous TNG simulation analyses (\citealt{Genel2018, Pillepich2019}). 
Note that in \citet{Baldry2012} the major-axis radius is adopted, and we used the correction introduced by \citet{Roy2018} to convert these radii to circularized radii. 
Looking at Fig.~\ref{fig_R}, we find an overall good agreement of the size--mass relationship of our galaxies with the ones from various observational works. 

In particular, the blue/star-forming galaxies are compatible with the size--mass relationship from \citet{Shen2003}, \citet{Baldry2012} and \citet{Roy2018}, within the scatter. 
Also, we have a good agreement with previous {TNG100-1 simulation} analysis from \citet{Genel2018}. 
{A reason for this latter agreement is that part of this relationship depends on the intrinsic properties of the galaxies, namely the sizes.
Here, our inferences based on mock images of simulations, before the convolution with the PSF, do not differ much from the standard analyses (e.g. in \citealt{Genel2018}).} 
{ On the other hand, we find a significant disagreement with the relationship of blue/star-forming galaxies from the TNG50 simulation (\citealt{Pillepich2019}) except for stellar mass $\sim 10^9\rm M_\odot$. The TNG50 results disagree also with TNG100-1 from \citet{Genel2018}, even though theseauthors adopt a similar galaxy definition based on subfinder and same physics. 
The basic remaining differences are the redshift, as TNG50 is available only for $z=0.5$, and the spatial resolution (i.e. $R=0.288\kpch$ of TNG50 versus $R=0.74\kpch$ of TNG100-1). 
For the former, we could expect that, given the effect of the size growth with redshift (see e.g. \citealt{Trujillo2007, Roy2018}), the measured sizes at $z=0.5$ are an underestimate of those at $z=0-0.2$, and hence the observed offset could be larger if the predictions at these lower redshifts were available. 
For the latter, the resolution making galaxies systematically larger could be the better detailed subgrid physics governing the feedback (stellar feedback, galaxy winds, AGN etc.), hence making feedback more efficient at redistributing the baryons over larger scales.
At the very small masses, Log$\rm M_{\rm star}/\rm M_\odot<9$, the absence of the plateau shown by the TNG50 star-forming galaxies is due to the better resolution that does not make the galaxy sizes to saturate around the spatial resolution of TNG100-1.} 

Looking at the red/quenched galaxies, we find a very good agreement with \citet{Baldry2012}, {while our results  deviate slightly from \citet{Roy2018}, which also disagree with \citet{Baldry2012},} at high and low masses. 
On the other hand the results of \citet{Shen2003} seem to strongly disagree with all other inferences. 
{A reason why these latter results deviate from all observations and simulations could be the galaxy selection, as these authors use the S\'ersic-index} index to separate red early-type from blue late-type galaxies, while our selection, as well as the one from \citet{Baldry2012}, makes use of the colour--mass diagram to separate red/quenched from blue/star-forming galaxies.
Our size--mass relationship seems also to deviate from the one of the {TNG100-1 simulation} from \citet{Genel2018} at stellar masses $\lesssim10^{9.8}{\rm M}_\odot$, but it is hard to ascertain the reasons of this, {except for a simulation resolution effect. 
Unfortunately, we cannot check this in higher-resolution TNG simulations (e.g. TNG50; \citealt{Pillepich2019}) as these authors did not provide the size-mass relationship for low redshift passive galaxies}.
We can just remark that this plateau (i.e. the flat distribution at the low-mass end in both panels in Fig. 4) does not have observational support.

%---------------figure_galaxyL_M---------------------------%
\begin{figure*}
\centering
\includegraphics[width = 0.99\textwidth]{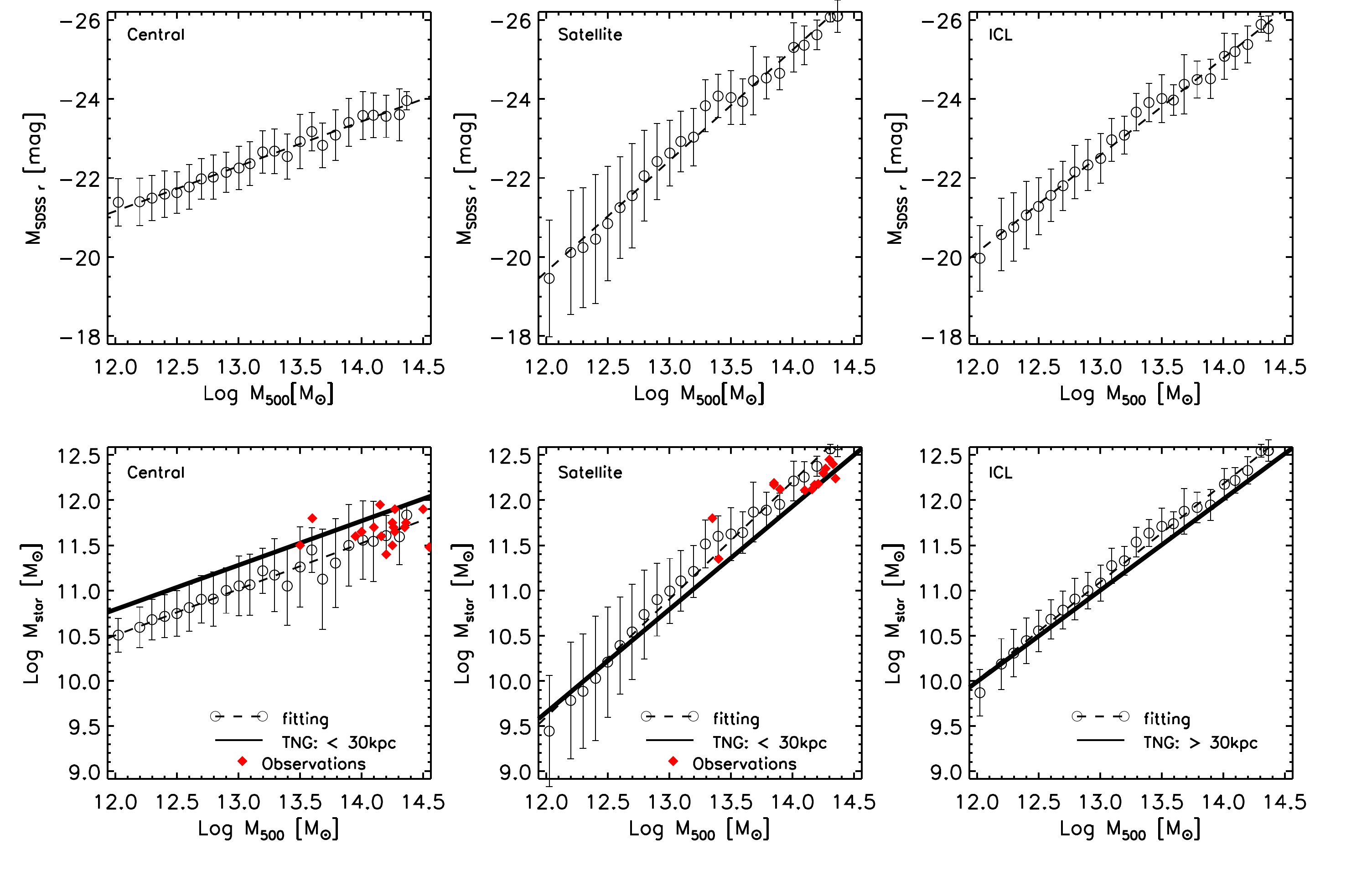}
\caption{
Dark matter halo mass--luminosity ({\it top row}) and stellar mass ({\it bottom row}) relationships for centrals ({\it left-hand panels}) , satellites ({\it middle panels}) and ICL component ({\it right-hand panels}), from the mock imaging. 
Data points with error bars and dashed lines are median values at given dark matter halo mass bin and the linear fit for each component, respectively.
Error bars represents the standard deviations. 
The solid lines in three bottom panels are the results from \citet{Pillepich2018b}.
The red {diamonds} are the results from observations \citep[][]{Gonzalez2013,Kravtsov2018}.
}
\label{fig_L_M}
\end{figure*}
%-------------------------------------------------------%

{Finally, we can discuss in detail the possible reasons of the other discrepancies with observations.}
{First}, we need to recall that the sizes from \citet{Baldry2012}, have been converted from the major-axis size, and this conversion depends on the constraints of the axis ratio in observations. 
This is possibly more difficult to constraint for the blue/star-forming galaxies, which are intrinsically flatter and more strongly affected by the seeing, which tends to make objects rounder. 
In general, an underestimate of $b/a$ would cause an overestimate of their corrected circularized radii, which is possibly the reason of the systematic larger sizes for the \citet{Baldry2012} blue galaxies.  
Another source of biases could come from our size definition (from the growth curve), which is different from {the 2D S\'ersic fitting of the surface-brightness profiles generally used for observed galaxies.} 
As demonstrated in \citet{Trujillo2001b}, sizes obtained via growth curve are underestimated with respect to the S\'ersic model ones for larger S\'ersic-index, {particularly for} larger luminosity/mass galaxies. 
This could be the case for our disagreement with \citet{Roy2018} in the high-mass end.
Hence, we expect that using a S\'ersic model approach to our mock observations could possibly alleviate this discrepancy, but this is beyond the purpose of this paper and will be investigated in future analyses.
{Finally, despite the fact that we find} a rather good agreement with \citet{Baldry2012} at almost all masses, our estimates below $\lesssim 10^{9.4}\rm M_\odot$ are consistent with spatial resolution of the {TNG100-1 simulation} and are less secure. 
Indeed, the apparent plateau can be due to some resolution effect, which would also explain the deviation from \citet{Roy2018}.

Moving to previous IllustrisTNG studies, we want to compare our approach in detail with the standard simulation analyses. 
\citet{Genel2018} studied the scaling relationships of galaxy samples defined by the subfinder { at $z=0.1$}. 
In particular, they derived the observational galaxy size relationship with galaxy stellar mass at ${\rm M}_{\rm star} >10^{10.5}\Msun$, with a scatter $\sim 0.1 $ dex.
{They do reproduce the correct trend of the relationship}, with more massive galaxies having larger galaxy sizes, and blue galaxies being larger  than the red ones (solid lines in Fig.~\ref{fig_R}), {but} their results are generally underestimated with respect to our size estimates and those from \citet{Baldry2012}, except for the massive red galaxies. 
The offset is especially significant for the blue galaxies, although here they seem to be more consistent with the results of \citet{Shen2003}.
The differences between our results and those of \citet{Genel2018} could have two main causes.
The first obvious one is the galaxy definition. 
They use subfinder to define galaxy size, which is larger than that on our sample defined by mock images.
{The second is that the characterization of galaxy samples in \citet{Genel2018} is made according to the galaxy sSFR.} 
{We tested different galaxy selections in our sample and found that the sSFR has a very minor impact on the overall shape of the size--mass relationship for both blue and red galaxies, though.}

\subsection{Central galaxy, cluster satellite, and intra-cluster light (ICL)}\label{L_M}
In this final subsection, we briefly discuss the global properties of the galaxies obtained by the SBLSP in Section \ref{methods}, with respect to the intra-cluster light (ICL) of the cluster/groups in which the galaxies are found. 
This latter component represents, by definition of our segmentation procedure, all the stellar mass (and luminosity), that remains after the galaxy detection. 
Hence, an inappropriate choice of the threshold adopted to define galaxy boundaries would reflect on both the luminosity and mass distribution of galaxies and the final amount of ICL in a a given cluster.

{To check this, we} compared our results with independent estimates of galaxy stellar masses and luminosities in clusters, both in simulations and observations (when available), after having separated galaxies in `centrals', (i.e. the cluster dominant systems) and `satellites' (i.e. all other galaxies).    
In particular, we concentrate on the linear correlation between the luminosity (expressed as absolute magnitude in the SDSS r-band, ${\rm M}_{\rm SDSS r}$) and stellar mass (${\rm M}_{\rm star}$) of galaxies {and} the mass of their host dark matter halos \citep[e.g.][]{Leauthaud2012, Lin2012, Gonzalez2013, Pillepich2018b, Kravtsov2018}.

{In Fig.~\ref{fig_L_M}, we plot the stellar content of the different galaxy types (central galaxies in left-hand panel, and the cumulative mass of all satellite galaxies in middle panel) and the ICL (right-hand panel) against the dark matter halo mass of the host group/cluster.
This latter is given as virial mass at the radius, $R_{500}$, corresponding to a mean halo density being 500 time the critical density for universe. 
In the same figure, we also show the best linear fits to the luminosity--virial mass (top panels) and the stellar--virial mass (bottom panels) as dashed lines.
The best-fitting parameters are reported in Table~\ref{fitting_para}.}
We find that the stellar mass of the central galaxies scales as ${\rm M}_{\rm{star}}\propto {\rm M}_{500}^{\alpha}$ with the best-fitting slope of $\alpha=0.51\pm0.02$.
Similarly, the slope is $\alpha=0.45\pm0.03$ for the luminosity of central galaxies.
We also compare the stellar mass estimates of our central galaxies with previous IllustrisTNG analyses (black solid line in the right-hand bottom panel of Fig.~\ref{fig_L_M}, from \citealt{Pillepich2018b}) and observations (red {diamonds}, from \citealt{Gonzalez2013} and \citealt{Kravtsov2018}).
{The observational sample, in particular, covers a narrow mass range, from ${\rm M}_{500}\approx10^{13.5}\Msun$ to ${\rm M}_{500}\approx10^{15}\Msun$, where there are only 21 clusters.
This may represent a significant selection effect when comparing the trends of observations with simulations in what follows.}

In Table~\ref{fitting_para}, we find that the slope of our central galaxies, $0.51$, is slightly steeper than the one in \citet{Pillepich2018b}, of $0.49$.
Moreover, the stellar mass in our sample is $\sim 0.2-0.3 $ dex lower than the prediction in \citet{Pillepich2018b}.
We remark that the stellar mass of the observed sample of central galaxies
\citep[][]{Gonzalez2013, Kravtsov2018} is estimated within the central
$50 \kpc$, and show a flatter slope with respect to $\rm M_{500}$ than the one from \citet{Pillepich2018b} and the one from our mock galaxy observations, as reported in Table~\ref{fitting_para}. 
More significantly, though, \citet{Pillepich2018b} clearly overestimate the stellar mass being systematically greater than suggested by the relationship derived from the observations, despite the fact that hey use a smaller radius (30 kpc) to measure the enclosed mass.
In contrast, our data points and linear fit agree almost perfectly with most all the observed galaxies, especially at the massive end.

Satellite galaxies in our sample, in the middle bottom panel of Fig.~\ref{fig_L_M}, also show a relationship between the stellar mass and dark matter halo mass that has a slope, $a=1.30$, steeper that the one, $a=1.14$, in \citet{Pillepich2018b}.
Furermore, the intercept $b=12.21$ in the satellite stellar mass--halo mass relationship is higher than {that of $11.93$} in \citet{Pillepich2018b}.
Consequently, in our sample, dark halos with ${\rm M}_{500} \gtrsim 10^{13}\Msun$ contain more satellite galaxies than the one in \citet{Pillepich2018b}, and, vice versa, dark matter halos with ${\rm M}_{500} \lesssim 10^{13}\rm M_\odot$ contain fewer satellite galaxies.
As shown in this figure, the distribution of the observed galaxies is systematically above the best-fitting line in \citet{Pillepich2018b}, but in a fair agreement with our linear correlation at all masses.

{Looking at the slopes of the relationships in Table~\ref{fitting_para} of central or satellite galaxies, we see that our estimates are slightly steeper than those in \citet{Pillepich2018b} and that the observed sample is shallower that both simulation results. 
However, this latter difference could be a selection effect resulting from the narrow mass range and small sample, as discussed above.}

Finally we consider the same correlation in the ICL (right-hand column in Fig.~\ref{fig_L_M}), as the ICL is the major stellar component in clusters, and it affects the stellar mass estimation of central galaxies.
From the best-fitting parameters in Table~\ref{fitting_para}, we see that the slope and intercept of ICL stellar mass--halo mass relationship in \citet{Pillepich2018b} and in our mock observations are similar.
In particular, the ICL stellar mass-halo mass relationships both show that more massive halos contain more ICL.
{The ICL fraction covers a large range, from 20\% to 50\%, and with a median ratios, 27$\pm$14\% (luminosity), 24$\pm$13\% (stellar mass).
This result is consistent with observations \citep[e.g.][]{Murante2004, Budzynski2014, Burke2015, Contini2019}, despite the large scatter caused by different observed environment and ICL definitions.}
{However, looking at the bottom right-hand panel, the ICL stellar mass estimated by us is more massive than that in \citet{Pillepich2018b}, at almost all halo masses. }
This discrepancy is statistically interpreted by the scatters of fitting parameters, and physically caused by the different galaxy definitions.

{Overall, these direct comparisons of the data from observations and simulations have demonstrated that the discrepancies {between TNG predictions} and data are strongly attenuated if the observed quantities, specifically the stellar mass, are compared to observational-like quantities derived from the same simulations with our mock imaging method. 
This means that there is not a substantial failure of the physical model behind the simulations, but rather that the way in which the physical information is extracted from simulations can introduce some apparent biases.}

%---------table1-----------------------------------------%
\begin{table}
\centering
 \begin{tabular}{l l c c c}
 
    \hline
    relationship & definition & slope ($a$) & intercept ($b$) & scatter\\
    \hline
    &&&\\
    ${\rm M}_{\rm stars, cen} - \MF$& Obs & 0.33 & 12.24&0.17\\
    ${\rm M}_{\rm stars, cen} - \MF$& $< 30$ kpc & 0.49 & 11.77& 0.12\\
    ${\rm M}_{\rm stars, cen} - \MF$& mock  & $0.51$ & 11.52 & 0.32\\
    &&&\\
    ${\rm M}_{\rm stars, sat} - \MF$& Obs & 0.75&12.52&0.10\\
    ${\rm M}_{\rm stars, sat} - \MF$& $>10^8 \Msun$& 1.14&11.93&0.22\\
    ${\rm M}_{\rm stars, sat} -\MF $& mock  & $1.30$&12.21&0.38\\
    &&&\\ 
    ${\rm M}_{\rm stars, ICL} -\MF $& $>30$ kpc & $1.01$&12.01&0.13\\    
    ${\rm M}_{\rm stars, ICL} -\MF $& mock  & $1.08$&12.17&0.24\\
    \hline
    ${\rm L}_{\rm r, cen} - \MF$& mock  & $0.45$ & 11.23 & 0.45\\
    &&&\\
    ${\rm L}_{\rm r, sat} -\MF $& mock  & $1.12$ & 11.95 & 0.89\\ 
     &&&\\ 
    ${\rm L}_{\rm r, ICL} -\MF $& mock  & $0.98$&11.87&0.59\\   
    &&&\\
    \hline
    
\end{tabular}
\caption{ 
The stellar mass budget in TNG groups and cluster at $z=0$: best-fitting parameters to the relationships presented in Fig.~\ref{fig_L_M}. 
The adopted fitting functions are the same as in \citet{Pillepich2018b}: $y = a m +b$, where $m={\rm log}_{10}( \MF/ \Msun) -14$, $y = {\rm log}_{10}(\rm M_{\rm star} / \Msun)$ for the stellar mass, and $y = {\rm log}_{10}({\rm L}_{\rm r}/{\rm L}_{\odot})$ for the galaxy luminosity.
Note that {the y-axis} in the top panel of Fig.~\ref{fig_L_M} represent the absolute magnitudes of galaxies. 
{These are related to the galaxy luminosities through the equation ${\rm M}_{\rm r}=-2.5\log ({\rm L}_{\rm r}/{\rm L}_{\odot})+4.64$.
In the table, the observational results (labeled by Obs) are taken from \citet{Kravtsov2018}, combining with the data in \citet{Gonzalez2013}.
We also  show  the slopes of the main subhalos within $30\rm{kpc}$ ( $<30\rm{kpc}$) and subhalos with stellar mass more massive than $10^8 \Msun$ ($>10^8 \Msun$) from \citet{Pillepich2018b}, respectively.}
}
\label{fitting_para}
\end{table}
%-------------------------------------------------------%

%---------------figure_simulation---------------------------%
\begin{figure*}
    \centering
\includegraphics[width = 0.49\textwidth]{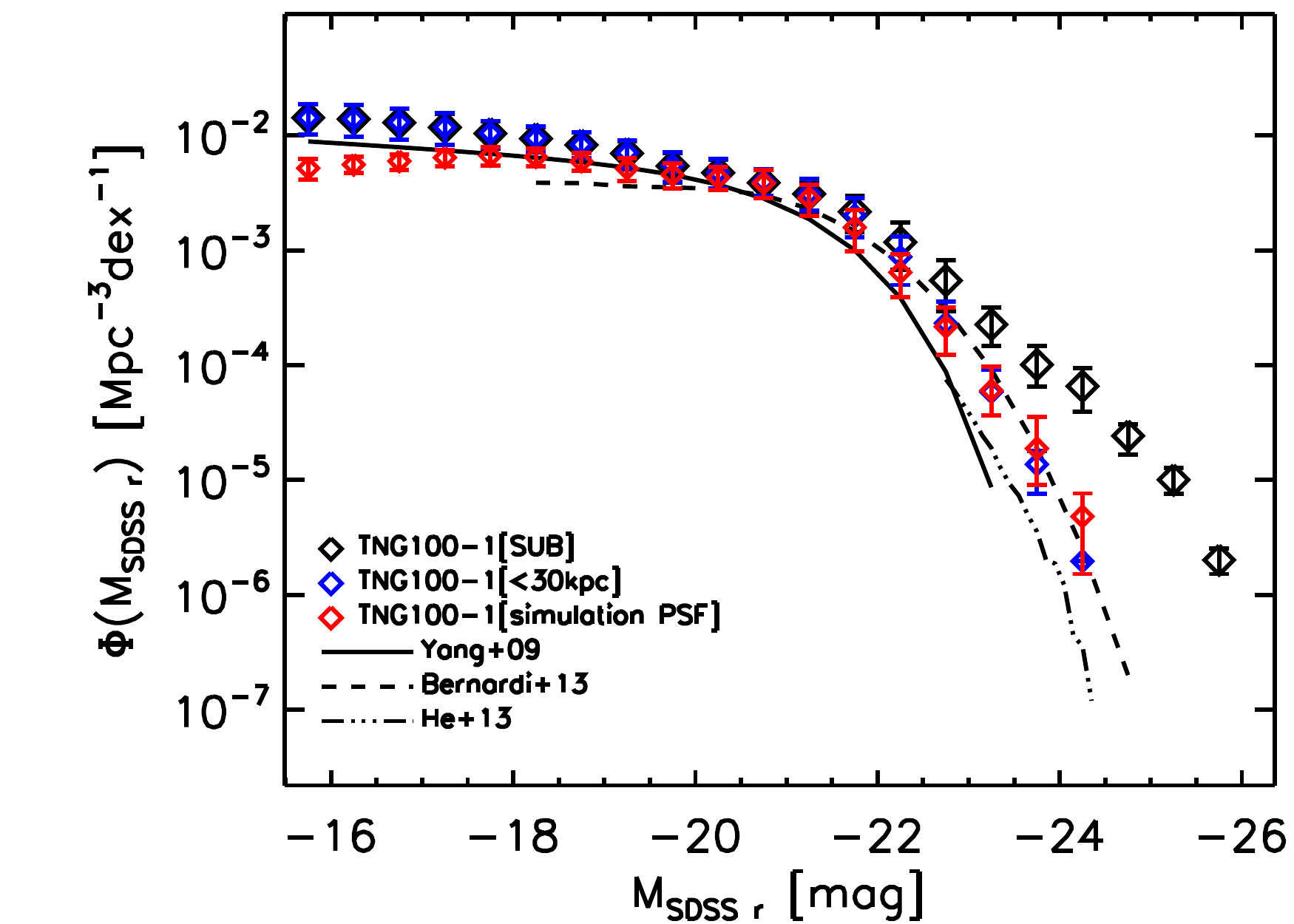}
\includegraphics[width = 0.49\textwidth]{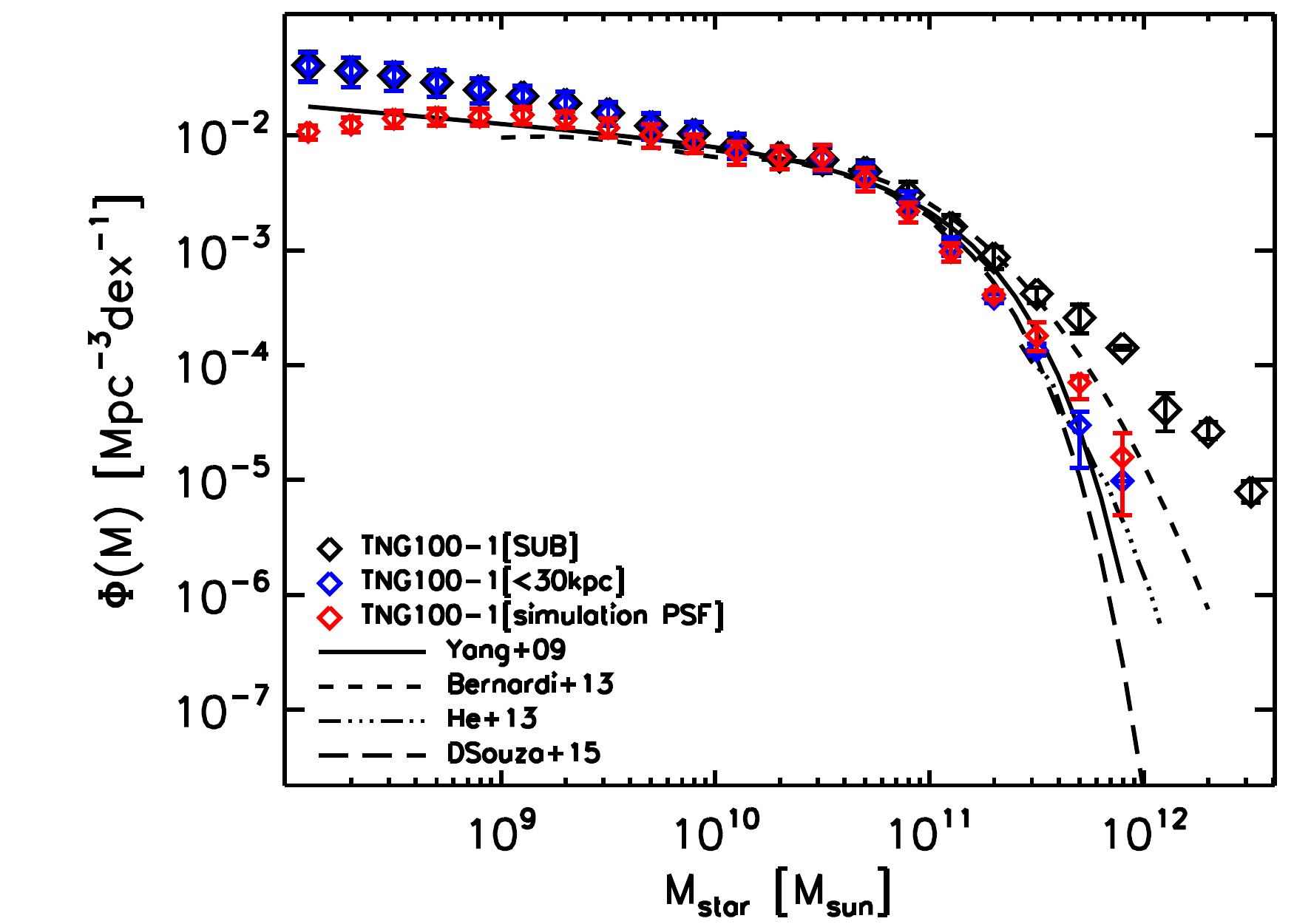}
\caption{{\it Left-hand panel}: the galaxy luminosity functions. {\it Right-hand panel}: the galaxy stellar mass functions. The black, blue and red {diamonds} are results for galaxy samples within the radius of subhalo, $30\kpc$, as well as obtained from our mock galaxy sample in {TNG100-1 simulation}, respectively. Our results in both panels are obtained by convolving the projected images with {\it simulation-PSF}, and selected by the criterion in Section \ref{population}.  The black lines are the observational results from SDSS in various papers (black solid line for \citet{Yang2009}, black dotted line for \citet{Bernardi2013}, dot-dashed line for \citet{He2013}, long-dashed line for \citet{D'Souza2015}, respectively), obtaining by their fitting formulae, expect that the results from \citet{He2013}. The fitting lines roughly cover the galaxy luminosity and stellar mass ranges in each set of observations.  The error bars for the galaxy sample defined with the radius of subfinder subhalos ([SUB]) and $30\kpc$($[<30\kpc]$) distance to shubhalo center are  calculated by averaging the results at three redshifts ($z=0.01,\ 0.1,\ 0.2$). Meanwhile, the error bars for our mock galaxy sample are calculated by averaging the results of nine galaxy samples (three projecting planes, xy, yz, zx, for each redshift).}
  \label{fig_simulation}
\end{figure*}
%-------------------------------------------------------%
%---------------figure_obeffect----------------------------%
\begin{figure}
    \centering
\includegraphics[width = 0.45\textwidth]{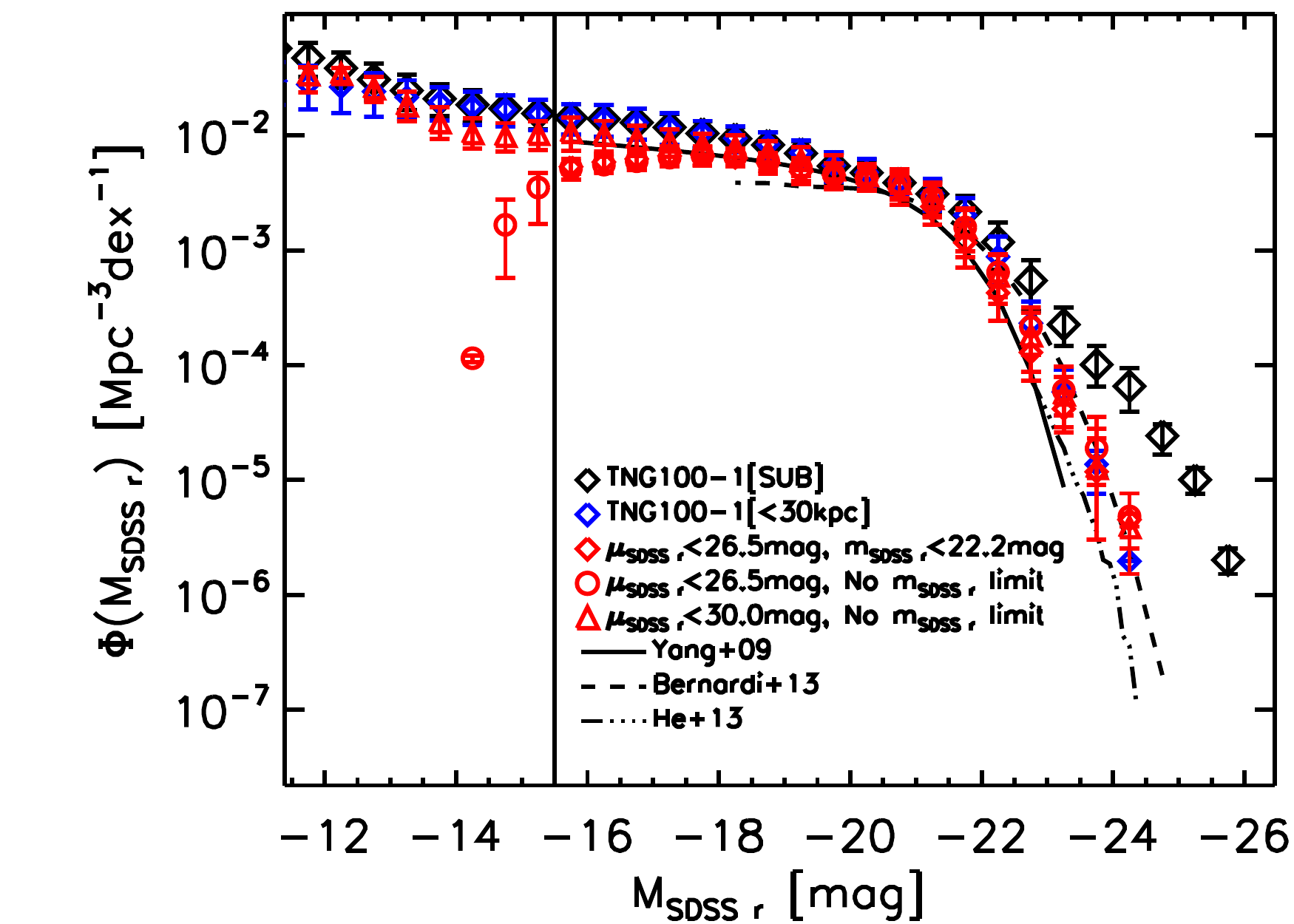}
\caption{This figure is similar with the left-hand panel of Fig.~\ref{fig_simulation}, but includes two observational parameter comparisons, as shown by the red triangle and circle symbols.
The vertical line denotes ${\rm M}_{\rm SDSS\ r}=-15.5 \ \rm mag$.
}
  \label{fig_obeffect}
\end{figure}
%-------------------------------------------------------%
%---------------figure_SDSS------------------------------%
\begin{figure*}
\begin{center}
\includegraphics[width = 0.49\textwidth]{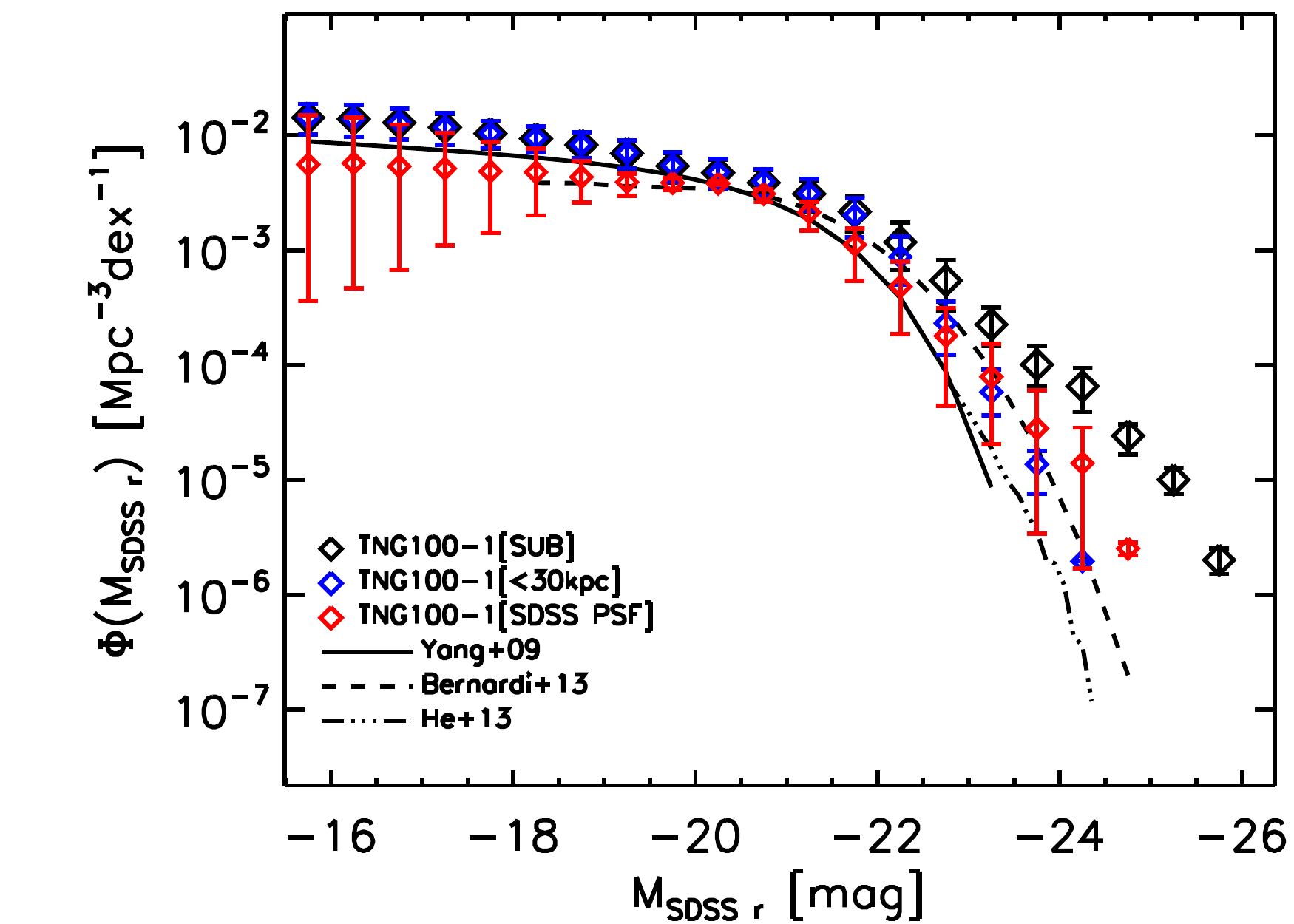}
\includegraphics[width = 0.49\textwidth]{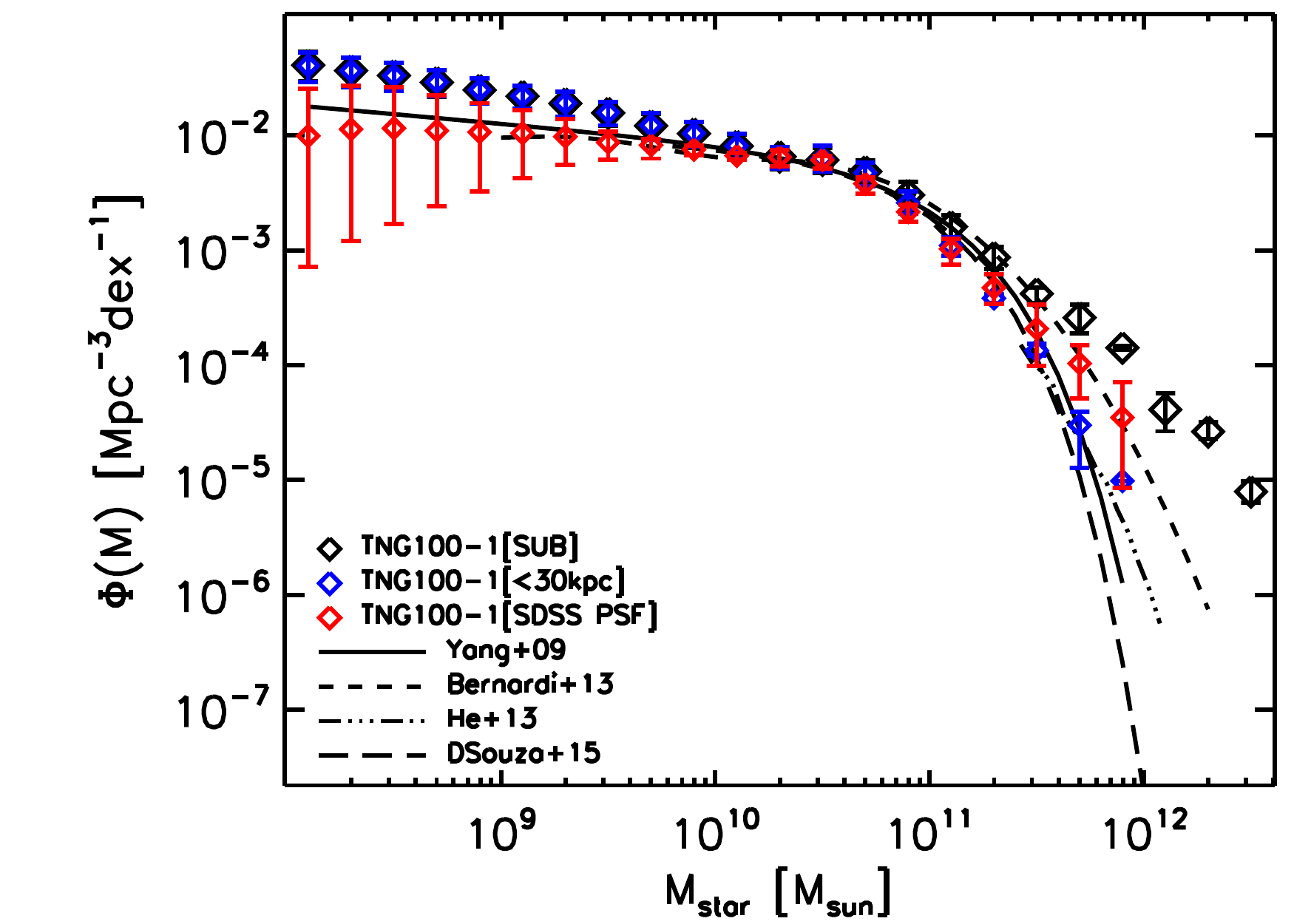}
\caption{
Same as Fig.~\ref{fig_simulation}, but it shows the GLF (left panel) and GSMF (right panel) obtained by convolved with {\it SDSS-PSF} ($\omega=1.43$) in TNG100-1 simulation.
}
\label{fig_SDSS}
\end{center}
\end{figure*}
%------------------------------------------------------%
%---------table2-----------------------------------------%
\begin{table*}
\centering
 \begin{tabular}{l c c}
    \hline
Symbol &  observational limit of surface brightness & observational limit of galaxy apparent magnitude\\ 
 & $\mu_{\rm SDSS\ r}$ & ${\rm m}_{\rm SDSS\ r}$\\   \hline
    &&\\
 
    red triangle      & 30 mag  & No limit\\
    red circle        & 26 mag  & No limit\\
    red {diamonds} & 26 mag  & 22.2 mag\\
    &&\\
    \hline
      
\end{tabular}
\caption{ The main observational parameters for data denoted by red symbols in Fig.~\ref{fig_obeffect}. 
Note that ${\rm M}_{\rm SDSS \ r}=22.2 \ \rm mag$ is approximate to ${\rm M}_{\rm SDSS \ r}=-15.5 \ \rm mag$.
}
\label{table_obeffect}
\end{table*}
%-------------------------------------------------------%

\section{Galaxy Luminosity And Stellar Mass Functions}\label{GLF_GSMF}
Having characterized the mock galaxy samples from simulation snapshots, we can now proceed to analyse their GLF and GSMF.
These are shown in Fig.~\ref{fig_simulation}, in left-hand and right-hand panels, respectively.  
In particular we also show: (1) the mock galaxy sample from the projected images convolved with SDSS-PSF in Fig.~\ref{fig_SDSS}; (2) the galaxy sample defined within the radius of subfinder subhalos (labelled as SUB) and (3) the same galaxy sample but with the mass estimated within a $R=30\kpc$ radius (labelled as $<30\kpc$).
We also illustrate some observation results from SDSS, as fitting formulae, e.g. from \citet{Yang2009, Bernardi2013, D'Souza2015}, or interpolated datapoints, for example from \citet{He2013}. 

\subsection{GLF and GSMF from mock observation catalogs and simulation-PSF}\label{GLF_GSMF_simulation}
As clearly shown in the left-hand panel of Fig.~\ref{fig_simulation}, the prediction obtained from our galaxy samples convoluted with simulation-PSF (red {diamonds}) is much lower than the GLF for galaxy sample defined by subfinder (black {diamonds}) at the bright end ($\rm {M_{SDSS \ r}}\lesssim -21.75$ mag), while the two predictions look similar at  intermediate luminosity ($-18$ mag to $-21.25 $ mag, including ${\rm L}_*$ galaxies). 
{At the faint end ($\rm{M_{SDSS \ r}}\gtrsim -18$ mag), though, our prediction shows an under-population of the GLF, with a slightly decreasing trend toward low luminosities, which is opposite to the increasing trend shown by subfinder-selected galaxies (black {diamonds}),} while the observational results \citep{Yang2009} fall in between.
{ The same effect is observed for the GSMF for galaxies with stellar mass $<10^9\rm M_\odot$, in that the GSMF in our mock observation sample is decreasing toward less massive systems, the one in subfinder-selected galaxies increase, and the one in the observed sample lies in between.}
This is a standard completeness effect driven by the observational set-up/conditions, mainly the limiting magnitude of the galaxy sample, {as we show for the luminosity function in Fig.~\ref{fig_obeffect}.} 
Here, we choose a series of different limiting surface brightness values, as in Table~\ref{table_obeffect}, to derive the corresponding GLF from the corresponding catalogs. 
As expected, the galaxy observational limit ${\rm M}_{\rm SDSS \ r}$ and surface-brightness limit $\mu_{\rm SDSS \ r}$ are the most important effects in our results. 
At the low magnitudes, the lower SB threshold {allows} us to identify more galaxies, and the faint end of the GLF becomes almost flat and closer and closer to the results predicted by applying the subfinder definition. 
Note that the faint end is close to the stellar mass resolution of simulations, and there could be a resolution effect that artificially reduces the abundance of small galaxies.
At high magnitudes, namely ${\rm M}_{\rm SDSS \ r}<-18.5$, the GLF is almost unchanged; that is, it is independent of the limiting magnitude assumed for the {SBLSP}.

This is interesting because the overall improvements introduced by our observational-like approach {(SBLSP)} are more evident at the bright end of both the GLF and the GSMF, which are much closer to the observed ones comparing with those based on subfinder. 
This is easily explained by the fact that our method allows us to more efficiently separate the galaxies in the center of the clusters from the non-negligible faint intra-cluster stellar component (\citealt{Tang2020}, see also Section \ref{methods}). 
The ICL is ubiquitous in the cluster centres \citep[e.g.][]{Burke2015, Zhang2019}. 
If included in the budget of the central galaxies and satellites, as shown in Figure 9 in \citet{Pillepich2018b} and Fig.~\ref{fig_L_M} in our paper, this ICL would cause an overestimate of their luminosity or stellar mass (see Fig.~\ref{fig_simulation}).

Furthermore, our results are close but slightly higher than the GLF and GSMF for galaxy sample defined within $30\kpc$ distance to subhalo centre  (blue {diamonds} in both panels) at the bright and massive ends.
On the other hand, the results at the faint end show that the GLF and GSMF for galaxy sample defined within $30\kpc$ are almost the same as the ones from galaxy sample defined from subhalos, and thus much higher than our predictions from mock observations.
Because the radius of massive BCG are typically larger than $30\kpc$, their luminosity or stellar mass are underestimated by the experiential radius cutting ( i.e.  $<30\kpc$), resulting in a lower GLF and GSMF at the bright end.
In contrast, at the low-mass end, galaxies have small radii, and thus the fiducial cutting-off of $30\kpc$ is large enough to include almost all the stellar particles in their host subhalos, resulting in the higher abundance than our predictions.

To check how realistic the different inferences from simulation are, it is useful to see how these compare with the GLF and GSMF from SDSS observations.
As shown in Fig.~\ref{fig_simulation}, overall, our prediction of the GLF and GSMF are located roughly in the range of observational results at the intermediate regime and the bright end.
For galaxies in the range of $\rm{M_{SDSS \ r}}$ between $-18$ mag and $-21 $ mag, our predicted GLF matches well with results of \citet{Yang2009} and \citet{Bernardi2013}. 
Here the measurements from the various IllustrisTNG analyses do not differ significantly either. 
For galaxies with $\rm {M_{SDSS \ r}}$ brighter than $-21$ mag, though, our `measured' GLF is well consistent with the results from \citet{Bernardi2013}, who carefully estimate the mass of the BCGs. 
Here, while the standard subhalo clearly deviates from observations, the $30\kpc$-selected sample seems to fit the observations fairly well, meaning that this {empirical fixed aperture} is indeed sufficient to capture the separation between the BCGs and the ICL. 
Similarly, looking at the GSMF, for galaxies in the mass range between $5 \times 10^8 \Msun$ and $7 \times 10^{10}\Msun$, our GSMF fits with results of \citet{Yang2009}, while for galaxies more massive than $7\times 10^{10}\Msun$, our predicted GSMF lies on top of the results from \citet{Bernardi2013}. 
Here we see that our results based on catalogs extracted by mock images are more accurate than those from other methods that deviate from the observed GSMF properly when BCGs are taken into account. 
This latter result can be partially due to the effect of the conversion between luminosities and mass in observations, as higher mass-to-light ratios might produce flatter slopes in the high-mass end, and lower mass-to-light ratios might produce steeper slopes. 
Overall, we find that the mock observation techniques we have proposed  seems to better capture the observed galaxy results, without introducing fine-tuning as for the {$30-\kpc$-aperture} approach.

Turning now to the faint end, the results from observed samples are almost flat, while for the mock galaxy samples we obtain a decreasing trend for ${\rm M_{SDSS r}}>-18$ and $\log \rm M_{star}<2\times 10^8 \rm M_\odot$, which deviates from observations of \citet{Yang2009} but rather similar to those of \citet{Bernardi2013}. 
In contrast, an increasing trend is found for the galaxy sample obtained from the standard subhalo or the one from the $30\kpc$ subhalo aperture, caused by a much higher galaxy number density distribution comparing with the observational results. 
This discrepancy between our estimate and standard IllustrisTNG analyses comes, again, from the different approach to the measurement of the galaxy parameters , which in our case is based on an observational approach. 
We note though that  {{\it this effect could be misinterpreted as an issue in the physical modeling in the simulation}, for example where the excess} of galaxies at the low end of the GLF or GSMF can be interpreted as an insufficient supernova feedback for quenching star formation in low luminosity/mass galaxies (e.g. \citealt{Genel2014}).

Overall, a major conclusion we can drive from this analysis is that {\it finding appropriate {strategies to reproduce realistic observational mocks of galaxy measurements} in simulations is an important factor to take into account when driving conclusion about the {missing physics or physical shortcomings therein}}.  
At the faint end, the upper boundary is the result for the  GLF and GSMF at redshift $z=0.01$, while the lower boundary is the result at redshift $z=0.2$, and vice versa at the bright end and the high mass end.
Note that the galaxy numbers at three redshifts are approximately equal. 
Other notice is that the low mass end is close to the mass resolution in {TNG100-1} simulation, and the observational effect might be one of the reasons, as shown in Fig.~\ref{fig_obeffect}.
Overall, the observational results at faint end is still uncertain.

%---------------figure_test--------------------------------%
\begin{figure*}
\centering
\includegraphics[width = 16cm, height= 7cm]{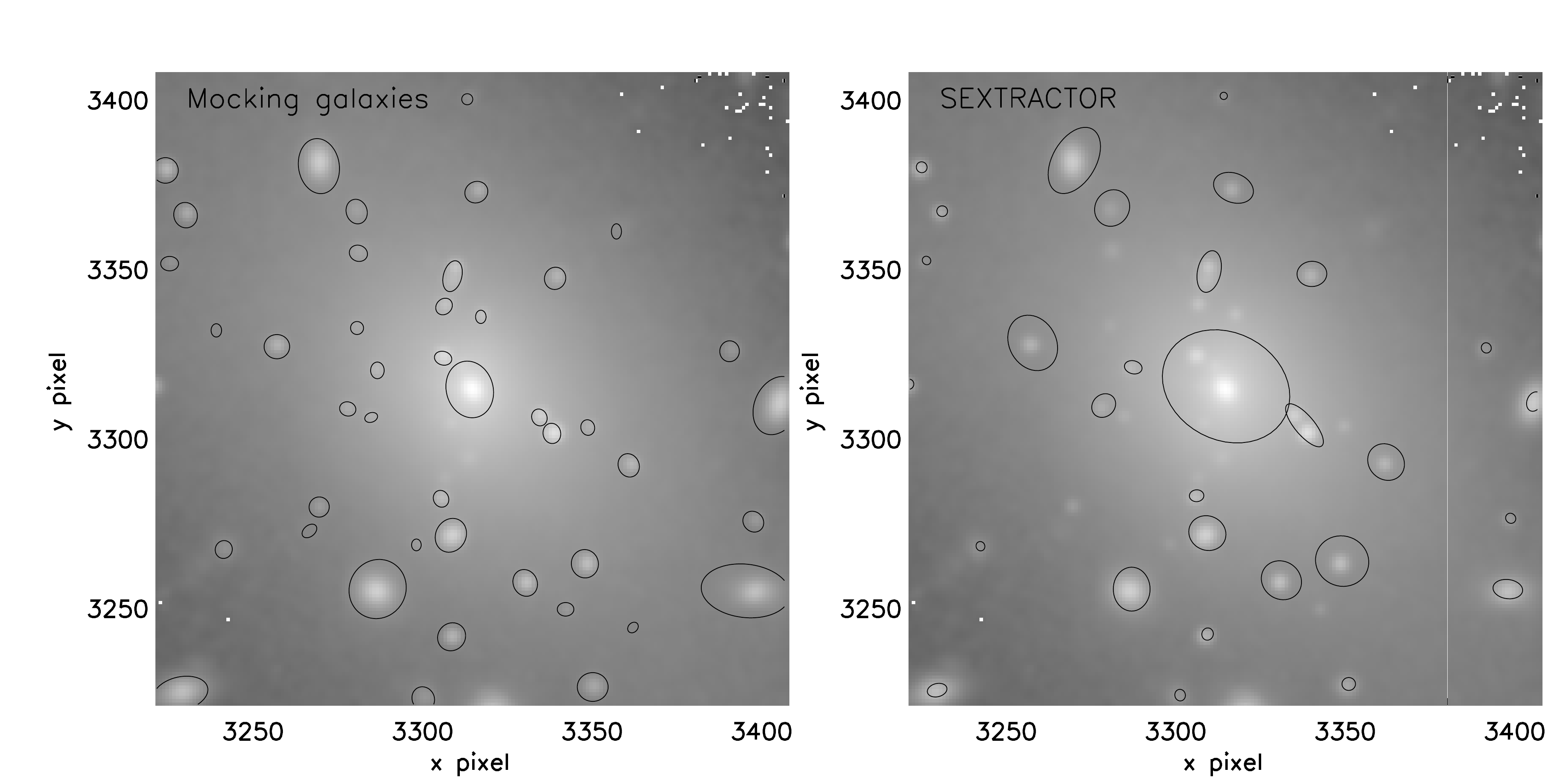}
\includegraphics[width = 16cm, height= 7cm]{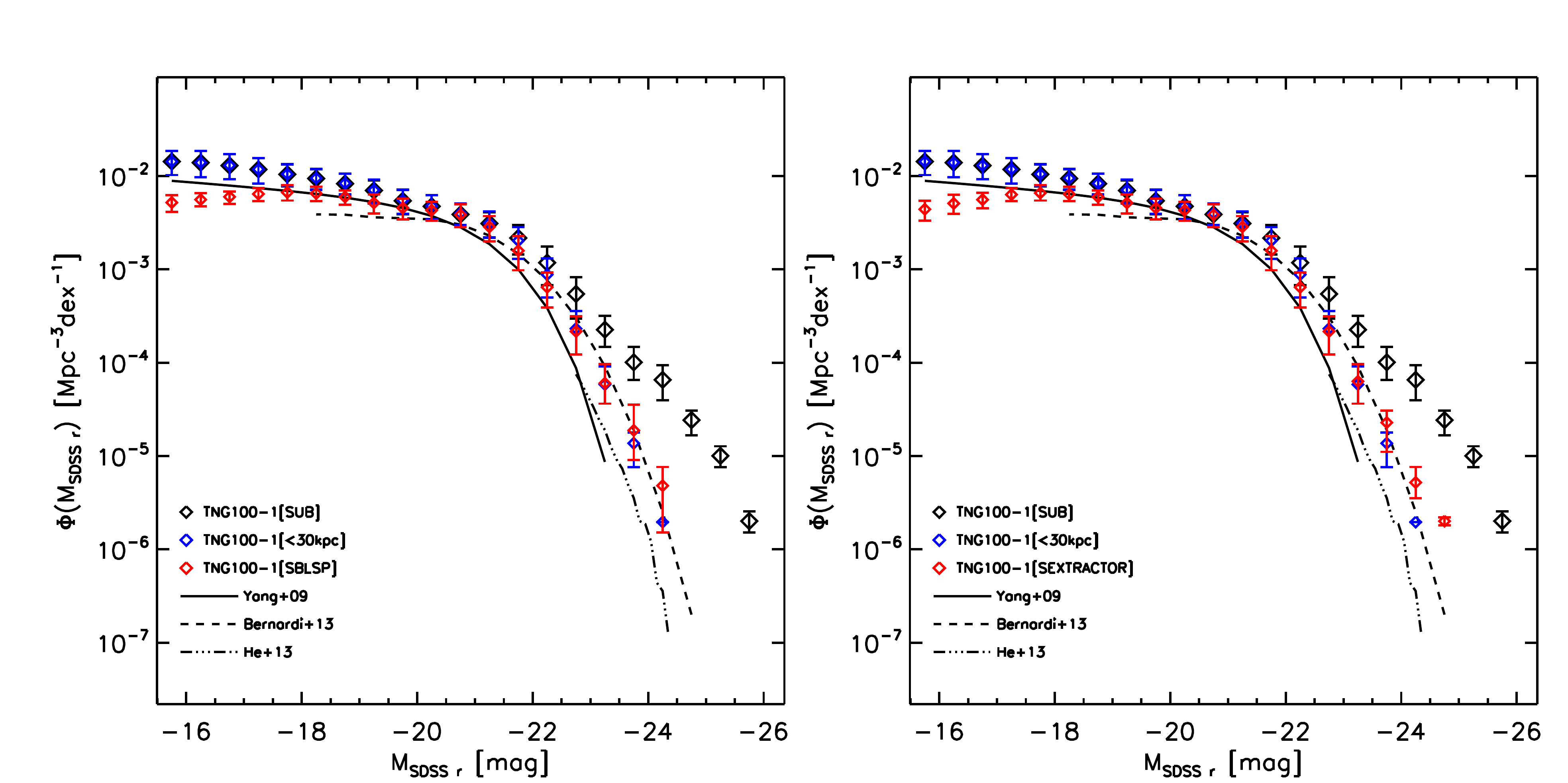}
\caption{{In the top row, the left-hand panel repersents the luminosity profile within the central region of the most massive group at $z=0$, with the galaxies defined by the SBLSP covered by the ellipses (this figure is the same as Fig.~\ref{fig_illustration}c); the  right-hand panelshows the luminosity profile within the same region in left-hand panel, but with the galaxies extracted by the SEXTRACTOR code.
Note that the colour-bar is as same as the one in Fig.~\ref{fig_illustration}.
The lengthen size of the side of the selected region is $135\kpc$, which is  the same as in the bottom row of Fig.~\ref{fig_illustration}.
The bottom row shows the GLFs (red {diamonds} in each panel) of galaxy sample in {TNG100-1} simulation defined by our SBLSP with simulation-PSF (left-hand panel, same with the left-hand panel of Fig.~\ref{fig_simulation}) and SEXTRACTOR code (right-hand panel). }
}
\label{fig_test}
\end{figure*}
%-------------------------------------------------------%

\subsection{GLF and GSMF with SDSS-PSF}\label{GLF_GSMF_SDSS}
To better mimic galaxy survey, a more realistic PSF should be adopted, rather than the theoretical one converted from simulation resolution.
The mock observation results using SDSS-PSF are shown in Fig.~\ref{fig_SDSS}.

To start, we should compare the difference of galaxy samples between  simulation-PSF and SDSS-PSF.
We simply check what the colour--mass diagram is look like with respect to the one with simulation-PSF convolution. 
It is found that the the parameters of van dan Bosch fitting line is $\alpha=$ 0.57, $\beta=$ 0.15 for SDSS-PSF, and $\alpha=$ 0.56, $\beta=$ 0.16 for simulation-PSF (also see Section \ref{color_sfr}).
Those two sets are almost same with each other.
The PSF-test of other galaxy properties also show a similar comparison above.
The most difference is the number of galaxies in two samples, as because of the much lager PSF width, there are fewer galaxies in SDSS-PSF, and fewer for higher redshifts. 

As can be seen, comparing with Fig.~\ref{fig_simulation}, our predictions for the GLF and GSMF become slightly higher for galaxies at the bright end or the massive end, while they become a bit higher and flatter at the faint end or the low mass end.
There is almost no significant change at the intermediate regime.
Because the SDSS-PSF is worse than the one obtained from simulation resolution, and thus it is more difficult to separate close components, the scatters are larger in Fig.~\ref{fig_SDSS} than in Fig.~\ref{fig_simulation}.
On the other hand, at the faint end, a worse PSF results inn fewer galaxies found at higher redshifts.
{We checked the galaxy numbers at three different redshifts for SDSS-PSF and simulation-PSF (not shown in figures)}. 
It is found that the galaxy number between two adjacent redshifts (i.e. between z=0.01 and z=0.1 or z=0.1 and z=0.2) differs by a factor of three for SDSS-PSF mock observations, while the difference of galaxy numbers at three redshifts is smaller than 10\% for simulation-PSF mock observations.
The redshift distribution of galaxy number for SDSS-PSF mocking is more close to the one in SDSS, particularly at the faint end.
Thus, the GLF and GSMF are almost flatt and more similar with the distribution in SDSS at the faint end or the low mass end, whereas they are slightly decrease using simulation-PSF, correspond to a closing equal galaxy numbers at three redshifts. 
Note that the upper boundary for $1 \sigma$ error-bars at the bright end corresponds to the lower boundary for $1 \sigma$ error-bars at the faint end.
It s also true for the GSMF.
Actually, similar to the results in Fig.~\ref{fig_simulation}, the behavior of GLF at the bright end and the faint end in Fig.~\ref{fig_SDSS}, as well as the GSMF at the massive end and the low mass end, looks like a seesaw, as it descends on the left-hand side while it arises at the right-hand side, and vise versa.

Considering the $1 \sigma$ uncertainty, our predictions for the GLF are consistent with observational results from \citet{Yang2009} at the faint end and with the observational results from \citet{Bernardi2013} at the bright end, except for the two or three brightest points.
The results are similar for the GSMF.
Please also note that both our predictions for the GSMF with simulation-PSF and SDSS-PSF are more close observational results than the corresponding predictions for the GLF.

These characteristics are effected from the PSF smoothing and low spatial resolution.
As discussion in \cite{Tang2018}, the PSF has an observable influence on the faint galaxies.
With a larger PSF width smoothing, the faint galaxies become darker, which results in fewer galaxies being observed at the faint end.
The massive galaxies are located in galaxy clusters, where the environment is complex. 
When smoothing with a PSF width larger than the pixel size, nearby components located in the cluster centre will inevitably be included by the massive galaxies.
Thus, many small details of structures in central region of massive galaxy clusters will be lost.
These influences are more obvious at higher redshifts.

{
\section{Systematics}\label{systematics}
In this section, we briefly discuss some possible source of systematics. We focus on three main aspects: (1) the source extraction method adopted; (2) the redshift dimming and $k-$correction; and (3) the cosmic variance.
}

\subsection{Testing galaxy segmentation against standard source extraction softwares}\label{test}
As discussed in Section \ref{methods}(iii), galaxies are identified on the projected image as luminosity peaks, and their edges are defined by different surface-brightness limits.
The SBLSP is similar to common source extraction tools used for real images. 
To check the consistency of our segmentation procedure with standard softwares for imaging data, we used one of the most popular source extraction code, SEXTRACTOR \citep{Bertin&Arnouts1996}.   
We accounted for the adopted observational set-up of our mock images in the SEXTRACTOR configuration file (e.g. pixel scale, seeing) and adopted different combinations of detection threshold and minimum detecting area to optimize the completeness in the source detection in our mock images.\footnote{ We stress here that these softwares have been optimized for real data and not for mock images, hence, despite the realism of these latter, we cannot be ensured that the straightforward application of these software produce satisfactory results in the source identification and photometry.} 

In the top row of Fig.~\ref{fig_test} we show one of the best optimized set-up, where we can see that SEXTRACTOR returns a number of sources that is almost equal to, if not smaller, than the one defined by our method.
Most of the losses in SEXTRACTOR are extended or faint galaxies.
For the most dense region, SEXTRACTOR cannot distinguish the galaxy that are closest to each other, {producing `blended' bigger detections.}
{We also found that the sizes of isolate galaxies extracted by SEXTRACTOR are generally smaller that the ones defined by our method.}
Overall, we can infer in the top row of Fig.~\ref{fig_test} that the galaxies defined by our method better match the density peaks. 
{We cannot exclude, though, that there might be more effective configuration set-ups that could improve the SEXTRACTOR results, but this further demonstrates that the optimizations of the standards tools for source detection often require accurate {\it fine-tuning}, while our method basically requires a single parameter (the SB threshold) to provide satisfactory results.} 
 
{To quantify the differences between the SBLSP and SEXTRACTOR, we show the GLFs derived from the catalogs obtained for both methods in the bottom row of Fig.~\ref{fig_test}.
Here, we see that the galaxy number density of the sample extracted by SEXTRACTOR is slightly lower than the one obtained with the SBLSP at the faint end ($\rm M_{\rm SDSS~r}>-17.5$), and viceversa at the bright end ($\rm M_{\rm SDSS~r}<-23$).
In particular, the slope of the decreasing trend toward the faint end is steeper for SEXTRACTOR than that for SBLSP.
Overall, the GLF of the SBSLSP is closer to the results of the observed sample, than the one of SEXTRACTOR. 
Once again, we cannot exclude the possibility that there might be some {\it ad hoc} set-up of SEXTRACTOR that could improve this result; however, this test shows that the use of software packages optimized for real data, cannot be easily extended to simulated data sets, and that in general the `galaxy photometry' is a potential source of bias in the comparison between simulations and observations.}

\subsection{Cosmological redshift effect}
{In our definition of observed magnitude, we have taken into account the effect of cosmological redshift dimming, which is given by the well known relationship $(1+z)^{-4}$. This is a minor effect for the low redshift samples we are considering, but it will become dominant for analyses that will include higher$-z$ simulation snapshot. 
However, as we have shown in \citet[][]{Tang2018}, this can still produce some effect on the measured GLF and  GSMF, if not accounted. In particular, these would result in higher number counts in the faint/low mass end, while, for $z<0.2$, it would produce almost no effect on the bright end. }

{On the other hand, an ingredient still missing in the generation of our mock images is the $k-$correction. 
This needs to be added in the definition of the observed magnitudes of the mock galaxies. 
Because we are currently limited to a relatively low redshift range (i.e. $z<0.2$), missing this correction can introduce a variation of $-0.1$ to 0.2 mag in $r-$band \citep[see e.g.][]{Westra2010, Chilingarian2010}, at the worse, which minimally impacts our final results.}

\subsection{Cosmic variance}
{Because of the difference between the cosmic volume of {TNG100-1} simulation and SDSS, we should take into account the effect of cosmic variance. 
Following \citet{Genel2014}, we assume that a cosmological box with a side length of $106.5\Mpc$ has a large enough volume so that overall galaxy statistics do not suffer from detrimental cosmic variance effects.}

\section{Conclusion}\label{conclusion}

{In this paper, we have investigated galaxy properties, for example colour, star formation rate, size, mass and luminosity distribution, galaxy mass and luminosity function of a mock galaxy sample from {TNG100-1 simulation of IllustrisTNG}.}
{Galaxy were identified using an improved surface-brightness-level segmentation procedure, dubbed SBLSP, based on an iterative surface-brightness level deblending in projected mock observations generated by converting the stellar particles from {TNG100-1} snapshots into seeing-convolved pixellized flux images.}
To obtain realistic observational conditions and reproduce realistic imaging-like data, {we considered a Moffat PSF with two FWHM definitions: one corresponding to the simulation softening length (the so-called simulation-PSF), and the other corresponding to the typical PSF from SDSS (the so-called SDSS-PSF).}
In addition, we comprehensively considered the observational limits and redshift range of SDSS in the mock image procedure.
Finally, we produced a series of galaxy samples with galaxy mass and luminosity range similar to SDSS observations.

{ The galaxy properties we obtained in our mock sample closely match those of real observed galaxies. 
In particular, we used the colour--mass diagram to separate the red sequence and blue cloud and showed a close match with previous SDSS results (e.g., \citealt{vandenBosch2008}). }   

{We have used these red/blue galaxies to study the properties of relevant scaling relationships like the size--mass relationship, in order to validate the realism of the mock galaxy sample and check the ability of these samples to return results closer to the real sample. 
We found that the effective radii derived by the mock galaxy catalogs derived with our SBLSP are in better agreement with the size--mass relationship of GAMA (e.g. \citealt{Baldry2012}) and KiDS (\citealt{Roy2018}), than the same quantities derived by simulated galaxy catalogs, based on standard group-finding algorithms (e.g. \citealt{Genel2018}, \citealt{Pillepich2019}). 
This is a first indication that use of observational-like approaches to the simulation predictions is critical to avoid biased conclusions about the discrepancies between observations and simulations.} 

{To further validate our mock galaxy sample, we studied the $\rm{M_{star}}$--$\rm{M_{halo}}$ correlation.
This is important in order to evaluate the fraction of stellar mass that has been defined by the SBLSP method (see Section \ref{methods}).
We have found that our mock central galaxies are less massive that those in \citet{Pillepich2018b}.
the distribution of the observed galaxies are systematically blow the best-fitting line in \citet{Pillepich2018b}, while they are in a fair agreement with our linear correlation at all masses.}

{We finally used the mock observational galaxy sample to study in detail the galaxy mass and luminosity function and compared them with the predictions for galaxy samples from simulations (using different apertures to define the galaxy samples) and observations from SDSS.}

{We found that, using the simulation-PSF to convolve the 2D surface brightness of the mock galaxies selected from simulations, our predicted GLF and GSMF are in better agreement with the ones obtained from observations (\citealt{Bernardi2013}) with respect to the predictions from the same TNG100-1 simulation but using galaxy samples defined by subfinder. 
On the other hand, using the SDSS-PSF, the predictions of GLF and GSMF from mock observations show a larger scatters, but are still consistent  with observations.
These results show that using observational realism (i.e. accounting for observational conditions in the data) plus mock observations of the data (i.e. appropriate galaxy segmentation) can reconcile some of the discrepancies between observations and simulations. 
This can affect the overall conclusions about the motivations of the observed discrepancies. 
For instance, an excess of bright/massive galaxies in simulations could indicate an insufficient AGN feedback (see e.g. \citealt{Weinberger2017}).
Vice versa, an excess of faint low/mass galaxies has been explained with insufficient supernova feedback (see e.g. \citealt{Pillepich2018a}), or even as a proof of a different DM flavour (e.g. \citealt{Lu2012}). 
Here we have shown that some of the tension between observation and simulations has to be traced to methodological approaches than theory.}

{Overall, this work has shown that finding appropriate strategies to reproduce realistic observational-like galaxy measurements in simulations is a key factor to develop further to derive robust comparisons between cosmological simulations and large surveys and fully understand what is the real physics still missing to fully explain galaxy and dark matter halo properties. }

%-------------------------------------------------------%

\section*{Acknowledgements}

The authors thank the anonymous referee for useful suggestions, the Illustris and IllustrisTNG projects for providing simulation and Drs, Weishan Zhou and Dandan Xu for useful discussions.
We acknowledge support from the National Key Program for Science and Technology Research and Development (2017YFB0203300) and the NSFC grant (No.12073089, No. 12003079).
L.T is also supported by the Fundamental Research Funds for the Central Universities, Sun Yat-sen University (No. 20lgpy176).
Most calculations are done on the Kunlun HPC in SPA, SYSU.
%-------------------------------------------------------%
\section*{Software}
This work is relied on IDL software, including \texttt{IDLAstro} and \texttt{Coyote Graphics Routines}.
The mock observation code may be made available upon reasonable request to the corresponding author.

\section*{ data availability}
IllustrisTNG simulations: \url{https://www.tng-project.org}
The mock galaxy samples or statistical data may be made available upon reasonable request to the corresponding author.
%-------------------------------------------------------%

\bibliographystyle{mnras}
\bibliography{LF_MF}%-------------------------------------------------------%

\label{lastpage}
\end{document}